\documentclass[aps,prl,preprint,superscriptaddress]{revtex4}

\usepackage{amssymb}
\usepackage{amsmath}
\usepackage{graphicx}
\usepackage{epstopdf}
\usepackage{color}

\def \diff{\ensuremath{\mbox{d}}}
\begin{document}

\title{Modelling and analysis of turbulent datasets using ARMA processes}

\author{Davide Faranda}
\email{davide.faranda@cea.fr}
\affiliation{
Laboratoire SPHYNX, Service de Physique de l'Etat Condens\'e, DSM,
CEA Saclay, CNRS URA 2464, 91191 Gif-sur-Yvette, France
}
\author{Flavio Maria Emanuele Pons}%
\affiliation{
Dipartimento di Scienze Statistiche, Universit\'a di Bologna,
Via delle Belle Arti 41, 40126, Bologna, Italy
}
\author{B{\'e}reng{\`e}re Dubrulle}%
\affiliation{
Laboratoire SPHYNX, Service de Physique de l'Etat Condens\'e, DSM,
CEA Saclay, CNRS URA 2464, 91191 Gif-sur-Yvette, France
}

\author{Fran{\c{c}}ois Daviaud}
\affiliation{
Laboratoire SPHYNX, Service de Physique de l'Etat Condens\'e, DSM,
CEA Saclay, CNRS URA 2464, 91191 Gif-sur-Yvette, France
}
\author{Brice Saint-Michel}
\affiliation{Institut de Recherche sur les Ph\'enom\`enes Hors Equilibre 
Technopole de Chateau Gombert 
49 rue Fr\'ed\'eric Joliot Curie B.P. 146 13 384 Marseille, France}
\author{\'Eric Herbert}
\affiliation{ Universit\'e Paris Diderot - LIED - UMR 8236
Laboratoire Interdisciplinaire des \'Energies de Demain - Paris, France}
\author{Pierre-Philippe Cortet}
\affiliation{Laboratoire FAST - CNRS, Université Paris-Sud, France}


\date{\today}

\begin{abstract}
We introduce a novel way to extract information from turbulent datasets by applying an ARMA statistical analysis. Such analysis goes well beyond the analysis of the mean flow and of the fluctuations and links the behavior of the recorded time series to a discrete version of a stochastic differential equation which is able to describe the correlation structure in the dataset. We introduce a new index $\Upsilon$ that measures the difference between the resulting analysis and the Obukhov model of turbulence, the simplest stochastic model reproducing both Richardson law and the Kolmogorov spectrum.
We test the method on  datasets measured in a von K\'arm\'an swirling flow experiment. We found that the ARMA analysis is well correlated with spatial structures of the flow, and can discriminate between two different flows with comparable mean velocities, obtained by changing the forcing. Moreover, we show that the $\Upsilon$ is highest in regions where shear layer vortices are present, thereby establishing a link between  deviations from the Kolmogorov model and coherent structures. These deviations are consistent with the ones observed by computing the Hurst exponents for the same time series. We show that some salient features of the analysis are preserved when considering global instead of local observables. Finally we analyze   flow configurations with multistability features where the ARMA technique is efficient in discriminating different stability branches of the system. 
\end{abstract}


\maketitle

\section{Introduction}
For a long time, experimentally testable predictions of turbulence properties have been influenced by available measurements. For example, hotwire velocity measurements have motivated statistical analysis of turbulent spectra or velocity increments computation, allowing the evaluation of Kolomogorov direct or refined similarity hypothesis \cite{kolmogorov1941local}. More recently, new sophisticated instruments and acquisition techniques, such as the Particle Image Velocimetry (PIV)  and the Laser Doppler velocimetry (LDV), have made  possible to measure instantaneous velocity fields with resolution equivalent to that of Large Eddy Simulations     \cite{kato1996laser,adrian2010particle,goldstein1996fluid}. With these  high quality datasets, it is now possible  to reconstruct the large scale flow dynamics and compute global observables even in relatively complex geometries such as in the non-homogenous, non-isotropic von K\'arm\'an flow \cite{ravelet2008supercritical,cortet2009normalized}. As more spatial and temporal scales are  becoming accessible to measurements, it is    important to extract all the possible information from a statistical analysis of the data as it may eventually lead to  test  new theoretical predictions.  In that respect, the integrated information obtained by measuring spectral features may not be sufficient to distinguish the contributions of different scales.  Moreover, filter response functions  used in spectral analysis may introduce spurious effects on the fast scales  hiding an intricate structure  \cite{zhou1993interacting}. Since present accessible measurements now give access to a large range of spatial scales, it therefore  seems  more promising to focus on turbulence properties in the physical space.\\  

Alternative statistical description of turbulence in the physical space actually date back to Kolmogorov and Obukhov \cite{kolmogorov1941local,obukhov1941distribution} and motivated formulation of stochastic models for the time evolution of turbulent observables. A now classical example is the Lagrangian stochastic model for the velocity of a passive tracer proposed by Thomson  \cite{thomson-jfm-1987}. In that model, the inertial range of Lagrangian turbulent velocity is described through a Langevin equation, involving parameters that are determined via the so-called {\itshape Well Mixed Condition} (WMC). This model is in fact equivalent to an autoregressive process of order 1, usually denoted AR(1) or ARMA($1,0$) (see below). Recent experiments however indicate that this simple model does not work for the velocity increments, which cannot be described by a simple standard Brownian motion as suggested by Obukhov \cite{frisch1996turbulence}.  Indeed, non normal corrections  originate from long correlations due to the intermittent character of turbulent flows. 
There are several models that suggest a more refined description, based, e.g, on the Rapid Distorsion Theory \cite{laval2003langevin}, on the account of the two-point two-time Eulerian acceleration-acceleration correlation \cite{friedrich2003statistics}, on temporal memory kernel \cite{baule2005joint}.  These approaches lead  to excellent approximations of the experimentally determined velocity \textit{pdf}'s, although an analytic solution for the model is still not available (for a review see \cite{aringazin2004stochastic,aringazin2004one}). However, it is not clear whether these models directly correspond to the features really observed in turbulent experiments or, in other words, how far is a real experiment from the theoretical idealization.  \\

To answer this question, as well as optimizing the information available from experimental measurements, it is mandatory to consider a more refined statistical analysis, able to account for  temporal memory effects as well as velocity dependent diffusion coefficients. A good candidate is given by analysis in terms of ARMA(p,q) processes, that have already been used to study problems ranging from geophysics to social science and finance  \cite{franses2000non,guiot1986arma}.  This analysis aims to represent the statistical properties of a time series $X_t$ using a model in which the value at time $t$ is a combination of the $p$ previous observations of the series - the so called auto-regressive part AR($p$) - and $q$ noise terms - the moving average part MA($q$) - with $p$ and $q$ chosen to be the lowest order to describe the series (see below). We observe that ARMA($p,q$) processes are also good candidates to describe  turbulent experimental data, since high $p$ orders correspond to high temporal memory and high $q$ orders correspond to a complicated structure of the diffusion coefficients. In the present paper,  we will apply the ARMA modeling technique to large datasets obtained in the (inhomogenous, anisotropic) von K\'arm\'an flow to illustrate the potential of this method.\\

The von K\'arm\'an experiment, in which the flow is generated in between two counter-rotating coaxial impellers, is a simple way to obtain experimentally a  large Reynolds number ($Re \sim 10^6$)  in a compact design \cite{marie2004experimental}. In the equatorial shear layer, fluctuations are large and exhibit similar local properties as in large Reynolds number experimental facilities devoted to homogeneous turbulence. Away from the shear layer, one observes a decrease of the turbulence intensity. Overall, the flow is strongly turbulent, so that the instantaneous velocity fields, measured by means of a PIV system, strongly differ in a non-trivial manner from their time average \cite{cortet2009normalized}. Although significant advancements in understanding the physics of this system  by statistical analysis \cite{cortet2010experimental} and from statistical mechanics approaches \cite{naso2010statistical} have been made,  several features of the flow remain unexplained and require further investigations. These include the nature of the phase transition recently discovered in the fully turbulent regime \cite{cortet2010experimental}, the forcing dependent stability of steady states \cite{briceprl} or the asymmetry of the torque probability distribution in different forcing conditions \cite{ leprovost2004stochastic,titon2003statistics}.
 These features are based on both local measurements (such as velocity measurements using PIV or LDV techniques) or global measurements, such as total angular momentum, energy or torque applied to the rotating disks by the turbulence (drag friction).
 For any of these local and global measurements,  we will define the ARMA($p,q$) model which better represents the  data, keeping in mind that the simplest model explaining the Kolmogorov turbulent spectra is the ARMA(1,0) model (see below). We then try to answer the following questions:

\begin{itemize}
\item How far is  a  von K\'arm\'an flow velocity time series  from an  ARMA($1,0$) model ?
\item  Can inhomogeneous anisotropic turbulence be  better described by  other ARMA($p,q$) models and for which orders?
\item  Is there a spatial organization of ARMA($p,q$) reflecting the spatial distribution of velocity inhomogeneities?
\item  Do different flow configurations correspond to different ARMA($p,q$) models?
\item  At a pure statistical level, is there an amount of information that the ARMA modeling can extract with respect to other techniques?
\end{itemize}

The main achievement of the paper is to suggest that these questions can be positively answered with a rather simple analysis. Moreover, once the order $p,q$ is identified, one has immediately a criterion to build    continuous stochastic models similar to the ones introduced in \cite{laval2003langevin,leprovost2004stochastic} for the quantities  analyzed. Our aim is thus to define a general technique which can be then used to analyze and critically extract information from any turbulence experiment.  In the present paper we underline the general procedure, leaving specific applications to future publications.
The paper is organized as follows: first we give an overview to present the relevance of the Obukhov model, then describe ARMA($p,q$) models for turbulence by giving a survey of their statistical and mathematical properties. Then we present the experimental set up and the quantities analyzed with the algorithm. Finally, we present and discuss the results obtained, outlining perspectives for the analysis of general turbulent datasets.

\section{ARMA  models of turbulence}

\subsection{From the Obukhov model to an ARMA($1,0$) process}
The celebrated phenomenological theories of Kolmogorov and Obukhov \cite{kolmogorov1941local,obukhov1941distribution} aimed to represent the complex phenomena of turbulence with a simple stochastic model. Thomson \cite{thomson-jfm-1987} was able to show that, in the inertial subrange, passive tracer Lagrangian velocities can be modeled by a Langevin equation (or Ornstein-Uhlenbeck process) with known coefficients; when discretizing this equation for simulation purposes, one can formally write it as an autoregressive process of order 1, usually denoted AR$(1)$ or ARMA($1,0$).
In particular, in the unidimensional case, the evolution of  the velocity and of the position of a tracer particle  $(u,x)$ can be described by the stochastic differential equations:
\begin{eqnarray}\label{T87}
\diff u &=& a(x,u,t) \diff t + b(x,u,t)\diff W \\
\diff x &=& u \, \diff t
\end{eqnarray}
where $\diff W$ are the increments of a Wiener process. In the same paper the determination of the coefficients $a$ and $b$ is discussed and it is found that, in Gaussian homogeneous turbulence, $a = -\frac{u}{T_L}$, where $T_L$ is the Lagrangian decorrelation timescale, while $b=\sqrt{C_0 \epsilon}$,  where $C_0$ is a universal constant and $\epsilon$ is the mean kinetic energy dissipation rate. This can be written, as suggested in \cite{tennekes-1982}, in terms of macroscopic quantities:
\begin{equation}\label{eps}
\epsilon = \frac{2 \sigma_u^2}{C_0 T_L},
\end{equation}
where $\sigma_u^2$ is the fluid velocity variance (equal to the Eulerian variance) and can be seen as a measure of the turbulence intensity.
Once the coefficients are known, one can write a discrete version of the Langevin equation (\ref{T87}):
\begin{equation}\label{langdisc}
\Delta u = -\frac{u}{T_L} \Delta t + \sqrt{C_0 \epsilon} \Delta W.
\end{equation}
We are now considering a discrete-time stochastic difference equation, so we can use a discrete-time index $t \in \mathbb{Z}$ and, rearranging, eq. (\ref{langdisc}) reads:
\begin{equation}\label{langAR}
u_t = \left(1-\frac{\Delta t}{T_L}\right) u_{t-1} + \sqrt{C_0 \epsilon} \Delta W.
\end{equation}
Denoting $ \left(1-\frac{\Delta t}{T_L}\right) = \phi$, $\sqrt{C_0 \epsilon \Delta t} = \sigma$ and recalling that $\{\Delta W\}$ are the increments of a Wiener process,  the equation can be rewritten as follows:
\begin{equation}\label{AR1}
u_t = \phi u_{t-1} + \varepsilon_t,
\end{equation}
where $\{\varepsilon_t\}$ are independent variables, normally distributed. Eq. (\ref{AR1}) is the expression of an AR$(1)$ process. To show that it is the simplest physical model which agrees both with Richardson law and the inertial range scaling proposed by Kolmogorov, it is sufficient to note that in an AR$(1)$ process,  the expected values of the velocity and the position scale in time respectively as:
\begin{equation}\label{toto}
  E[u^2(t)] \sim t, \quad E[x^2(t)] \sim t^3.
  \end{equation}
The second property is the Richardson law. Then, defining $\delta u=\sqrt{E[u^2(t)]}$ and $\ell=\sqrt{E[x^2(t)]}$, we get from eq.(\ref{toto})  $\delta u \sim \ell^{1/3}$ which can be seen as an equivalent of the Kolmogorov scaling.

\subsection{Generalization: ARMA($p,q$) model for turbulence}
The ARMA($1,0$) leads to a Markovian evolution for the Lagrangian turbulent velocity, and is unable to describe the  intermittency or memory that have been shown to exist in real flows.  In most laboratory turbulent flows, available datasets are time series of values of a physical observable at a fixed point or obtained by tracking Lagrangian particles. In our case, time series are obtained at fixed points in space; in this work, no spatial velocity profiles are studied.
This historically motivated the shift of paradigm from \textit{space} velocity increments to \textit{time} velocity increments defined as $\delta u_\tau=u(t+\tau)-u(t)$ and motivated computations of the  time  structure function. Of course, in situations where measurements are made on the background of a strong mean velocity $U$, scale velocity increments and time velocity increments can be directly related through the Taylor hypothesis $\ell=U\tau$. In situations such that the fluctuations are of the same order than the mean flow, however, the Taylor hypothesis fails. A suggestion has been made by \cite{pinton} to then resort to a \textit{local Taylor Hypothesis}, in which $\ell=\int dt u(t)$ where $u$ is the local rms velocity. This is equivalent to consider a scale such that $\ell\sim \tau  \delta u_\tau$  and may be seen as equivalent to modifying the \textit{space} Kolmogorov refined hypothesis into a \textit{time} hypothesis.\\
A natural generalization to take into account these features is thus to consider higher order ARMA$(p,q)$ models, exhaustively treated, in example, in \cite{brockwell_etal-springer-1990}. A summary of useful notions about ARMA$(p,q)$ modeling is provided in Appendix.
An ARMA$(p,q)$ model corresponds to discrete time, stationary stochastic processes $\{ X_t \}$ such that, for all $t$:
\begin{equation}\label{ARMA1}
X_t = \sum_{i=1}^p \phi_i X_{t-i} +  \sum_{j=1}^q \vartheta_j \varepsilon_{t-j} + \varepsilon_t.
\end{equation}
 $\{\varepsilon_t \}$ is assumed to be a white noise of variance $\sigma^2$ and the polynomials $\phi(z) = 1 - \phi_1 z_{t-1} - \cdots - \phi_p z_{t-p}$ and $\vartheta(z) = 1 + \vartheta_1 z_{t-1} + \cdots + \vartheta_q z_{t-q}$, with $z \in \mathbb{C}$, have no common factors. Notice the white noise assumption is a very general condition  and $X(t)$ will be normally distributed,  resulting by a linear combination of  independent and identically distributed  random variables.\\

From a physical point of view eq. (\ref{ARMA1}) is the natural extension of the ARMA(1,0) model corresponding to the Obukhov model by introducing a temporal memory structure:   Intuitively, the autoregressive part of the process expresses a dependence of the value of the process at time $t$ on a linear combination of its own $p$ previous values, while the moving average component introduces, at time $t$, a linear dependence of the $q$ previous values of the noise term. The quantification of  memory effect in real turbulent flows will then be made through fits of the data by an ARMA($p,q$) model, and measurements of how \textit{far} this model is from the ARMA(1,0) model. For this, we first need to define the notion of \textit{best} ARMA($p,q$) fit, and then the notion of 
\textit{distance} between the ARMA(1,0) and a given ARMA($p,q$).

\subsection{Model selection and characterization via correlation analysis}

The main idea of time series analysis through ARMA models is to select the linear model that fits the data in the most parsimonious way, so that diagnosis of the nature of the generating process, forecasting or Monte Carlo simulations can be performed. Model selection is a non-trivial step of the procedure that can be addressed essentially by two means: through correlation analysis or through information based criteria, such as the Bayesian Information Criterion ($BIC$) described below. \

For very simple processes, one can get access to the time dependence structure through computation of the auto-correlation function (ACF) and of the partial autocorrelation function (PACF) formally defined in the appendix.
In particular, for a MA$(q)$ process, the theoretical ACF is characterized by $q$ non-zero peaks, while the PACF decays exponentially or as a damped trigonometric function; for AR($p$) processes the PACF is characterized by $p$ non-zero peaks while the ACF decays exponentially. Hence, this fact allows  to rule out or confirm the validity of a AR($p$) or MA($q$) hypothesis by a simple inspection of the ACF and PACF.

In the general ARMA($p,q$) case, the simple correlation analysis described just above is not insightful. The model choice and the parameters estimation can be assessed by using the procedure introduced in \cite{box_etal-springer-1970}, which also takes into account more complicated (such as integrated and seasonal) models:
\begin{enumerate}
\item preliminary analysis: the series is plotted in order to identify possible trends in mean and variance or periodic behaviors. Since here we deal with physically stationary processes, no trends are expected;
\item identification on the basis of the estimated ACF and PACF (or applying information criteria, such as the $BIC$);
\item estimation through maximum likelihood techniques;
\item diagnostic checking, that is, testing the estimated  sequence for residual correlations (and normality or other distributive hypotheses, if required).
\end{enumerate}

\begin{figure}
\includegraphics[width=0.5\textwidth]{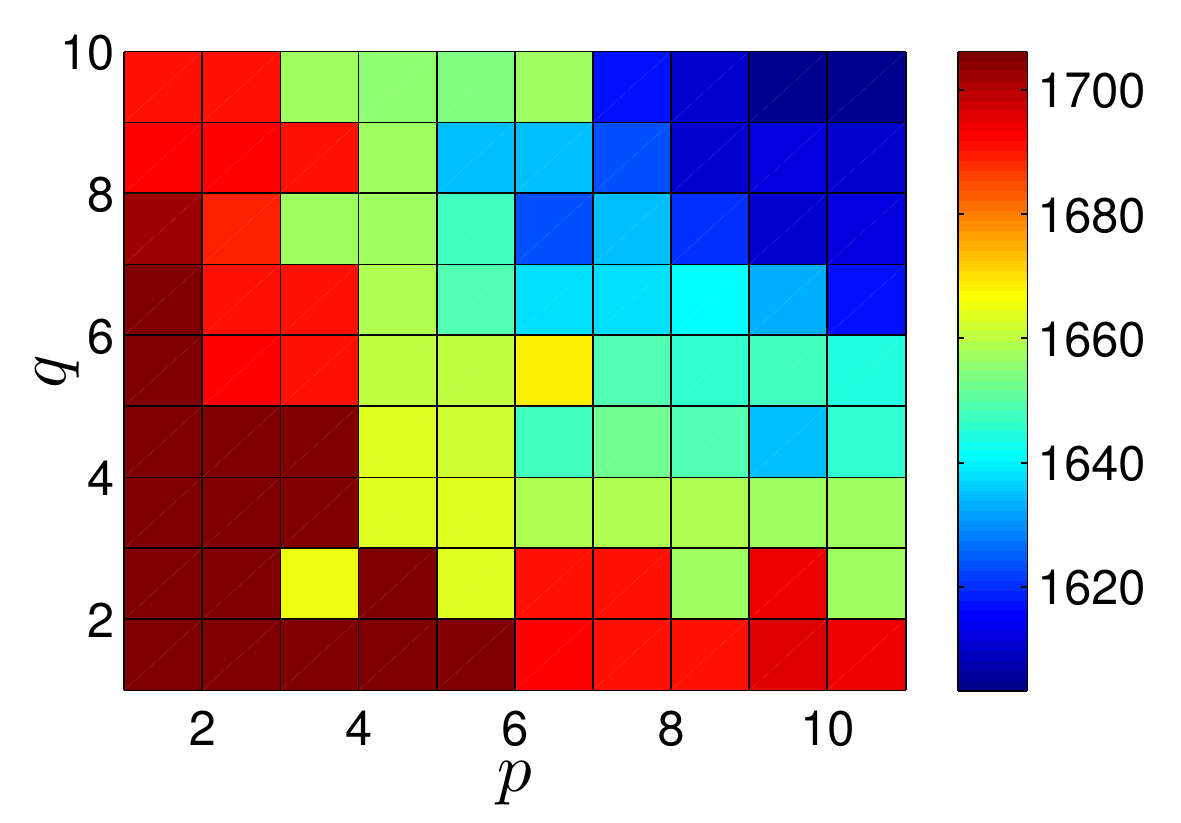}
\caption{ $BIC$ values resulting from different fits of ARMA($p$,$q$) model for a velocity time series consisting of 600 observations.} 
\label{BICtest}
\end{figure}

In the following analysis we perform the second step of the procedure fitting an ensemble of models with different ($p$, $q$) couples; we then choose the ARMA($p$, $q$) model with the lowest total order $p+q$ producing not correlated residuals. The serial independence of the residuals series is tested as described in Appendix A. As already mentioned, this phase could be based on the value of the $BIC$. In this case, the information criterion is computed for each model: the best fit is the minimum $BIC$ after the \textit{steepest descent}. The two methods provide the same results. 
First of all, we tested them on a synthetic time series of $10^5$ values simulated from an ARMA(3,1), obtaining a correct estimation of the model with both methods. To ensure that this technique is stable also for shorter time series,  in Fig. \ref{BICtest} we show a $BIC$ profile as a function of $p$ and $q$ for one of the analyzed velocity samples, consisting of 600 observations: both the methods lead to the choice of an ARMA(7,7). All the PIV time series have length $n=600$, while the typical LDV sample size is $n \sim 5 \cdot 10^5$.

\subsection{A measure of distance from Kolmogorov theory based on the Bayesian Information Criterion}
It is useful to concentrate the information obtained by analysis in a single index. We want to obtain a measure of the distance of the selected ARMA($p,q$) model from the ARMA($1,0$), namely the Thomson-Obukhov model. 

For a given dataset, the relative quality of a statistical model can be measured by the $BIC$, defined as: 

\begin{equation}
BIC = -2 \ln{\hat{L}} + k [\ln(n) + \ln(2 \pi)]
\end{equation}

where $\hat{L}$ is the likelihood function for the investigated model. For an exhaustive definition of this quantity, see \cite{casella1990statistical}. Since the likelihood function is maximized when the correct model is found, while goes to zero in case of mispecification, its logarithm grows for well-specified models, while diverges to $-\infty$ otherwise. Thus, the first term globally tends to become negative or to assume small values once the best model form is identified. On the other hand, the second term grows with the number of parameters times the sample size: so it serves as a penalization for the number of parameters, in order to avoid overfitting. In brief, when testing an ensemble of models for a certain dataset, the best one is identified by the minimum value of the BIC.

For a Gaussian ARMA($p,q$) model, it is expressed as follows:
\begin{eqnarray}\label{BIC}
BIC(n, \hat{\sigma}^2, p,q) & = & (n-p-q)\ln \left[ \frac{n\hat{\sigma}^2}{n-p-q} \right] + n(1 + \ln\sqrt{2\pi}) +\\
       & + & (p+q)\ln\left[ \frac{\left(\sum_{t=1}^n X_t^2 -n\hat{\sigma}^2  \right)}{p+q} \right]. \nonumber
\end{eqnarray}


Notice that $n$ is fixed by the experiment. The sample variance $\hat{\sigma}^2$ is computed from the sample and is a series-specific quantity. Thus, in order to obtain a meaningful definition of the distance from Kolmogorov model, the $BIC$($n, \hat{\sigma}^2, p,q$) must be normalized with respect to the Obukhov case $BIC$($n, \hat{\sigma}^2, 1,0$).

\begin{equation}\label{Intermittency}
\Upsilon= 1 - \exp \left \{| BIC(n, \hat{\sigma}^2, p+1,q) - BIC(n, \hat{\sigma}^2, 1,0) |\right \}/n,\quad 0\le \Upsilon\le 1
\end{equation}

This quantity tends to zero if the dataset is well described by the Obukhov model and tends to one in the opposite case.  We introduce the $p+1$ correction to magnify small $\Upsilon$ values.

\section{Experimental set-up and data processing}

In order to illustrate and apply these concepts, we have
worked with a specific axisymmetric turbulent flow: the
von K\'arm\'an  flow generated by two counter-rotating impellers
in a cylindrical vessel. The experimental set-up is described in \cite{briceprl,saint2013zero}. Here, we consider a configuration where the disks are exactly counter-rotating at frequency $f_1=f_2=F$, in the  two forcing
conditions associated with the concave (resp. convex)
face of the blades going forward, denoted in the sequel by
sense ($-$) (resp.  ($+$) ). The resulting mean velocity fields are quite similar, with two toric recirculations seperated by a mean shear layer (see fig \ref{meanvel}).  The forcing conditions however strongly influence the level of fluctuations, which are much higher in the ($-$) case.
The working fluid is water, with viscosity $\nu = 1.0 \times 10^{-6}$ m$^{2}\cdot$s$^{-1}$.  
The Reynolds number is defined as
$$Re = 2\pi F R^2\nu^{-1}$$ where $R$ is the cylinder radius. We introduce a cylindrical system of coordinates $\vec{x}=(R,\varphi,Z)$ with its origin at the center of the cylinder and the $z$-axis aligned with the impeller's rotation axis (see Fig.~1 of \cite{cortet2011susceptibility}). 

In the sequel, we analyze both local and global observables. As local observable, we will consider the time series of the modulus of the velocity fields $$ \vec{V} (\vec{x}, t) = [u(\vec{x}, t), v(\vec{x}, t),w(\vec{x}, t)]$$ obtained by PIV measurements.  Here $u$ is the component in the PIV plane described in terms of $R$, the radial distance from the center of the cylinder;  $v$ in terms of  $Z$, the vertical distance from the center, and $w$ is the normal component to the PIV plane (the azimuthal velocity in cylindrical coordinates). For the comparisons between PIV and LDV measurements we will consider the normal component only $w$. We will also address two important aspects of statistical modeling of turbulence: the appearence of multifractal cascades, usually studied via the computation of the Hurst Exponents, and the role of phase randomization, which permit to isolate the effects of intermittency related only to the phase of the signals.

As global observables, we consider first the normalized kinetic energy introduced by Cortet et al \cite{cortet2009normalized}:
\begin{equation}
 \delta(t)=\frac{\langle V ^2(t)\rangle}{\langle \bar{V}^2 \rangle}.
\label{delta}
\end{equation}
Here the brackets indicate the spatial average, and the bar a time average. $\delta(t)$ represents the ratio of the total kinetic energy of the instantaneous flow to the total kinetic energy of the mean flow.  As a second global observable, we also consider the torque time series $C_1(t)$ and $C_2(t)$ experienced by the two motors.  The goal is  to compare which part of information about the   flow  is carried by observables built using the velocity fields (such as  $\delta(t)$) and which  is carried by dissipation measurements such as the torques. 

The impeller speed $F$ and the applied torques $C_1$ and $C_2$ are related to the average dissipation rate in the experiment, $\epsilon$, through the injected power $\mathcal{P}$. The typical kinetic energy in the experiment, $\langle \bar{V^2} \rangle$, can be directly computed from the PIV data. Knowing these global quantities, it is possible to obtain rough estimates of two typical quantities of the turbulent flow, the Kolmogorov typical length and time scales $\eta$ and $t_\eta$ and the Taylor typical scale $\lambda$, using the dimensional analysis inspired by~\cite{kolmogorov1941local} and the identities of~\cite{taylor1935statistical}. These relations are only valid for homogeneous and isotropic turbulence, which is not the case here : we will use them anyway to get rough estimates, presented in Table~\ref{tab:scales} for experiments conducted in water at $F = 5$~Hz, with curved blades and under various forcing conditions.

\begin{table}
	\begin{tabular}{c | c | c | c | c | c | c}
	Sense & Cells & $\epsilon$ (m$^2 \cdot $s$^{-3}$ ) & $\langle \bar{V^2} \rangle$ (m$^2 \cdot $s$^{-2}$) & $\eta$ (m) & $t_{\eta}$ (s) & $\lambda$ (m) \\
	\hline
	\hline
	$(-)$  & 1   &  28  & 7.8  & $1.4 \times 10^{-5}$  & $1.9 \times 10^{-4}$ & $2.0 \times 10^{-3}$ \\
	$(-)$  & 2   &  9.2 & 2.6  & $1.8 \times 10^{-5}$  & $3.3 \times 10^{-4}$ & $2.0 \times 10^{-3}$ \\
	$(+)$ & 2   &  2.5 & 1.3  & $2.5 \times 10^{-5}$  & $6.3 \times 10^{-4}$ & $2.8 \times 10^{-3}$ \\
	\end{tabular}
	\caption{Typical length scales of the flow, using only global average quantities and usual turbulent identities~\citep{taylor1935statistical}. Data are obtained with curved blades in both rotation senses, for $f_1 = f_2 = 5$~Hz, using water. Two turbulent states coexist in the $(-)$ sense due to hysteresis~\cite{ravelet2004multistability}, one with one recirculation cell and another one with two cells. The one-cell state is unstable in the $(+)$ direction for $f_1 = f_2$.}
	\label{tab:scales}
\end{table}
 
\section{Results}
We begin the analysis of the datasets by showing how the ARMA procedure, described in the previous section, works on two velocity series extracted at two different locations from the same PIV experiment at $Re=10^5$ and for the ($+$) sense of rotation.  The series $|\vec{V}(t)|=\sqrt{\vec{V}(t)^2}$ and their ACF and PACF are represented in  Fig. \ref{armaex}. They have been obtained by   sampling the data at 15 Hz. By analyzing the structure of the ACFs and the PACFs, one observes immediately that they do not consist of a small number of discrete peaks out of the confidence bands. This excludes the possibility that the series can be represented by pure AR($p$) or MA($q$) processes. Moreover, it is clear that a by-eye determination of the order ($p,q$) is not possible.

This result is consistent with the non-Markovian behavior used to describe the torque measurements via stochastic models in \cite{laval2003langevin,leprovost2004stochastic}.  By implementing our best fit procedure, we find that the best ARMA model to fit the data depends on the measurement points: the series on the left-hand side of Fig. \ref{armaex} is fitted by an ARMA(1,1) model whereas the other one by an ARMA(4,2) model. This is of course not surprising, because the von K\'arm\'an flow is highly inhomogeneous. In the remaining of this section, we analyse the relationships between the flow inhomogeneous spatial structure and the ARMA fit structure by mapping the ARMA parameters.

\subsection{Velocity fields}
\begin{figure}
\includegraphics[width=0.5\textwidth]{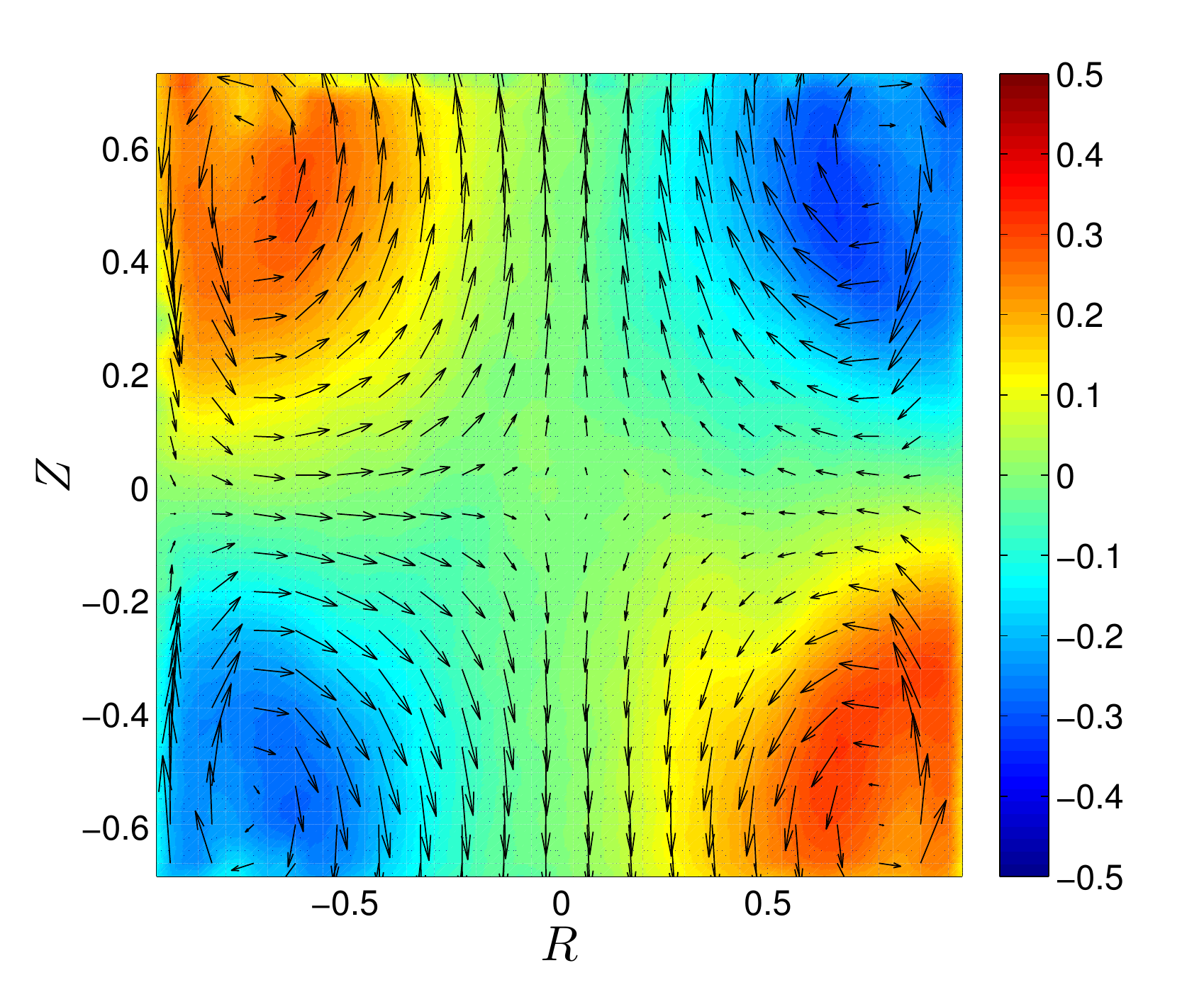}\includegraphics[width=0.5\textwidth]{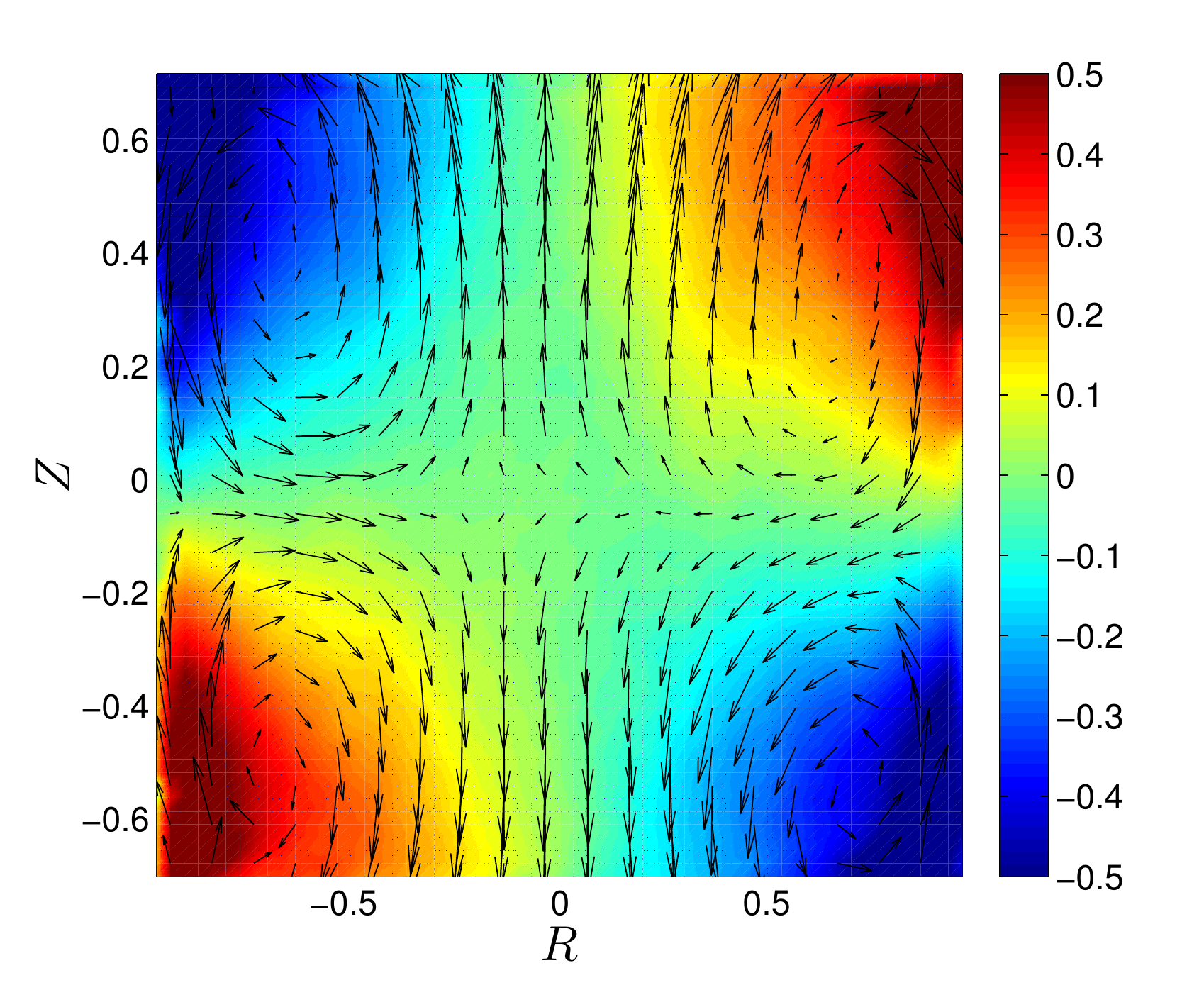}
\caption{ Structure of the mean velocity field for $\theta=0$. The arrows represents directions and intensities of the velocity components in the PIV plan averaged over time $(\bar{u},\bar{v})$.  The orthogonal component $\bar{w}$ is represented by the color scale. Left: ($+$) sense of rotation. Right: ($-$) sense of rotation.} 
\label{meanvel}
\end{figure}

\begin{figure}
\includegraphics[width=0.5\textwidth]{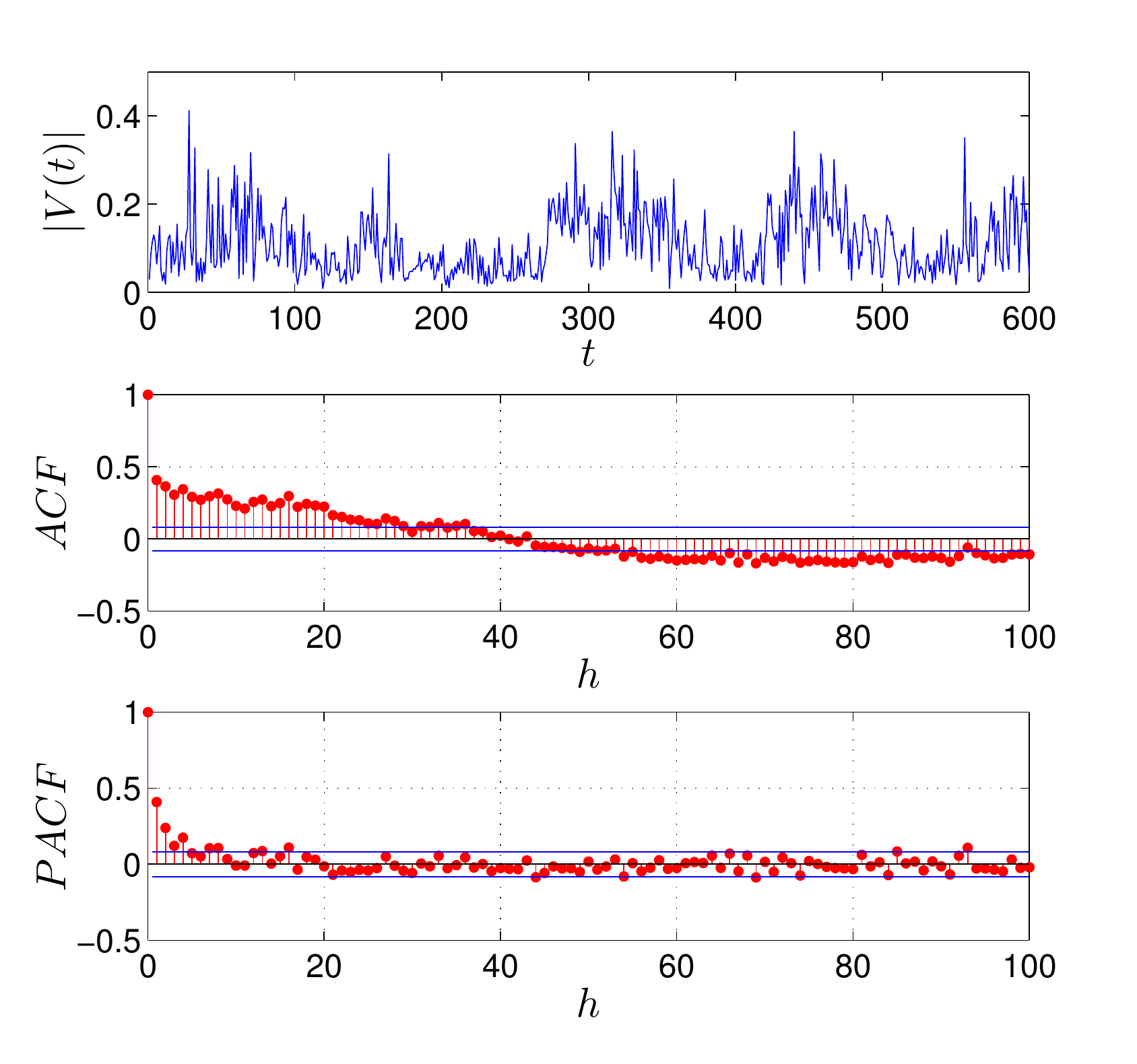}\includegraphics[width=0.5\textwidth]{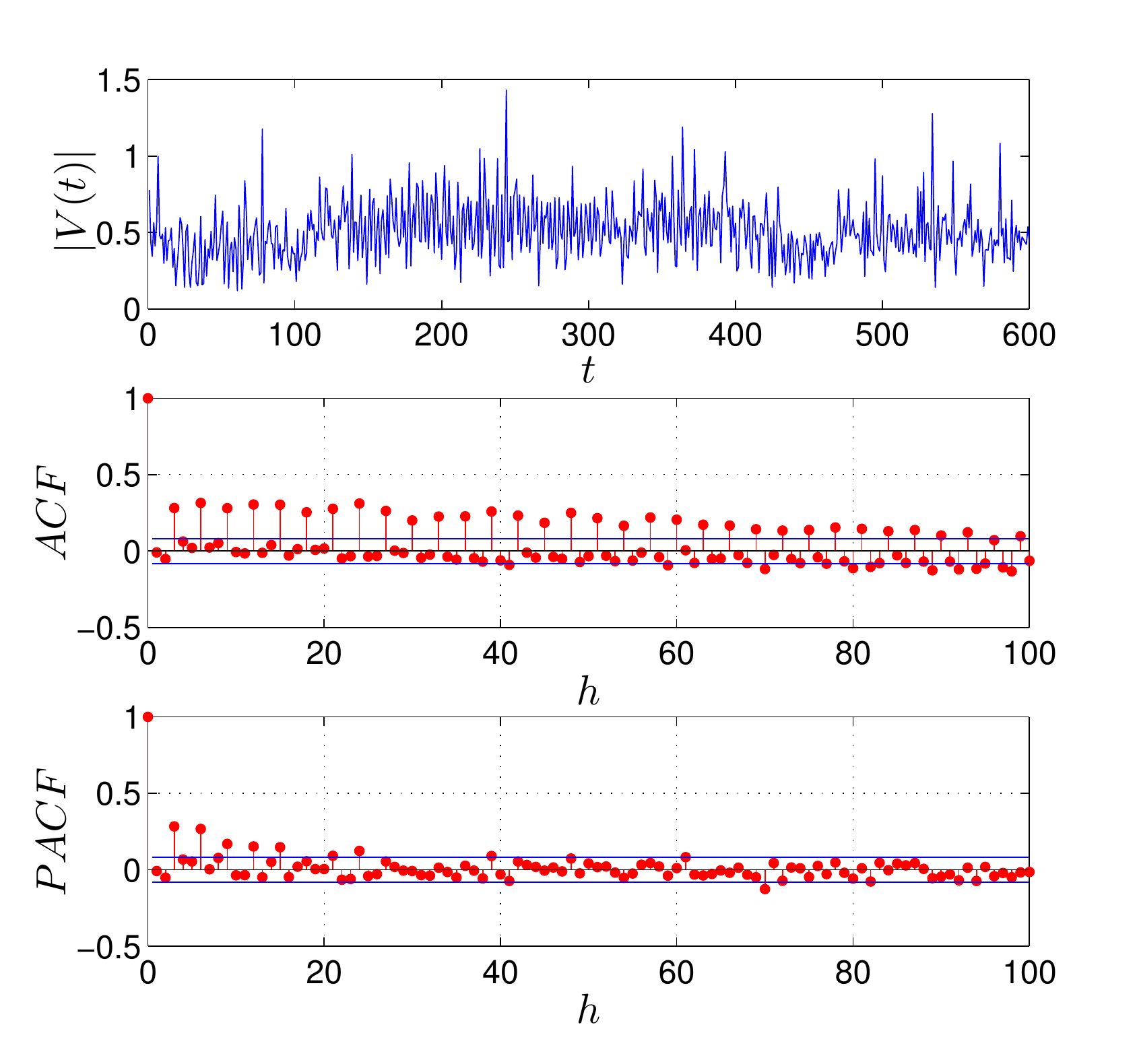}
\caption{ Two time series of $|V(t)|$ (upper panels) with their respective ACF functions (middle panels) and PACF (lower panels). $Re\simeq 10^5$, ($+$) sense of rotation. Blue lines in the ACFs and PACFs represent the confidence bands at the 95$\%$ confidence level. Sample frequency: 15 Hz. X-axis is in sample index.} 
\label{armaex}
\end{figure}

Let us now analyse the spatial structures obtained by applying the procedure described in the previous sections for a PIV field taken at $Re=2\cdot 10^5$, with mean velocity field provided in Fig. \ref{meanvel} for the ($+$) sense of rotation (left) and ($-$) (right).  The two pictures look extremely similar: one can immediately recognize the cells structure of the flows described in the previous sections.

\begin{figure}[H]
\includegraphics[width=1.0\textwidth]{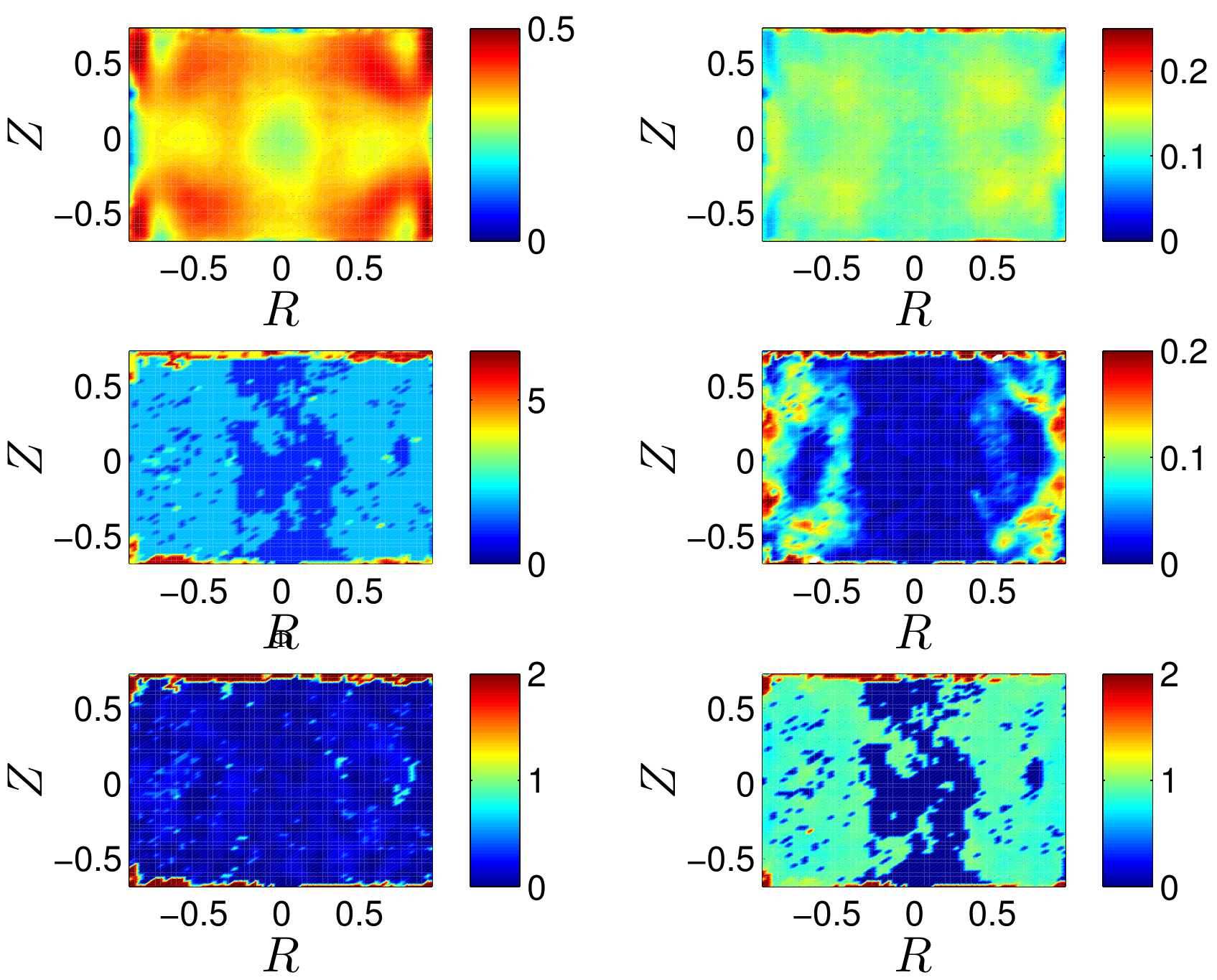}
\caption{ ARMA analysis for the ($+$) sense of rotation. Top left:  $|\bar{V}|$. Top right: $|\vec{V}(t)|$ standard deviation. Center left: total order $\mathcal{O}$ found  by fitting an ARMA($p$,$q$) to the $|\vec{V}(t)|$ data. Center right: distance from the Kolmogorov model $\Upsilon$ for the $|\vec{V}(t)|$ data. Bottom left: Sum of the  autoregressive coefficient $\Phi$. Bottom right: Sum of the moving average coefficients $\Theta$. } 
\label{armacon}
\end{figure}

\begin{figure}[H]
\includegraphics[width=1.0\textwidth]{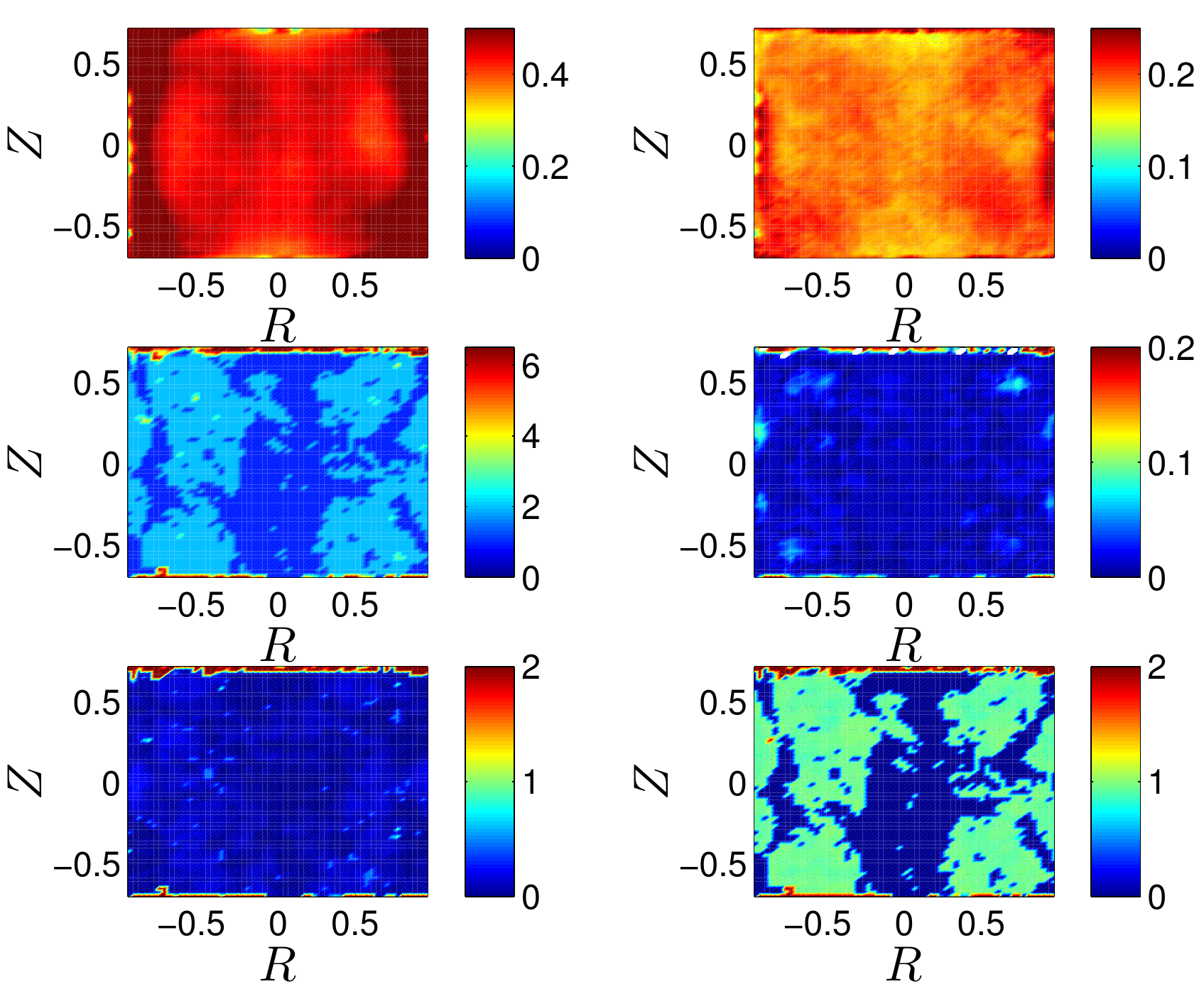}
\caption{ARMA analysis for the ($-$) sense of rotation. Top left: $|\bar{V}|$. Top right: $|\vec{V}(t)|$ standard deviation. Center left: total order $\mathcal{O}$ found  by fitting an ARMA($p$,$q$) to the $|\vec{V}(t)|$ data. Center right: $\Upsilon$ for the $|\vec{V}(t)|$ data. Bottom left: Sum of the  autoregressive coefficient $\Phi$. Bottom right: Sum of the moving average coefficients $Theta$. } 
\label{armaanti}
\end{figure}

 A full overview of the quantities computed by using the ARMA analysis is presented in Fig. \ref{armacon}  for the ($+$) sense of rotation and in  Fig. \ref{armaanti} for the ($-$) sense. 
Obviously, even if the four cells structure presented in Fig. \ref{meanvel} is recovered in both the situations, the average over time of $|\vec{V}(t)|$ denoted as $|\bar{V}|$ (top left panels)  and the standard deviation  of $|\vec{V}(t)|$(top right panels) show  remarkable differences between the two configurations. In the ($+$) sense of rotation the four cells structure is appreciable whereas higher mean values and fluctuations of $|\vec{V}(t)|$ are recorded in the proximity of the wall of the cylinder in the ($-$) rotation. In this latter configuration, the fluid is pushed to the side of the cylinder and higher turbulent fluctuations are registered, as described in \cite{ravelet2008supercritical}. Let us now analyze what happens to the quantities introduced by the ARMA analysis. 

\subsubsection{Order of ARMA}
We start by describing the behavior of the total order $\mathcal{O}=p+q$ of the processes fitted for each time series (middle left panels). In both senses of rotation, the highest orders are concentrated near the impellers. However, differences appear in the other regions of the domain. In the ($+$) sense of rotation, the highest orders are found in correspondence to the highest fluctuations. Near the center of the cylinder the orders are low and, for some of the series, the signal is indistinguishable from noise (order 0). In the ($-$) set up, the highest level of turbulent fluctuations contribute to homogenize the behavior in larger areas such that a weak four cells structure is recognizable.  This effect is probably linked to the presence of more homogeneous fluctuations in the flow (top right panel of Fig. \ref{armaanti}).   Even if the $\mathcal{O}$ color scale has been limited at $p+q=6$ for comparison with the ($+$) situation, we underline that much higher orders appear in the ($-$) setups near the impellers and, locally, at the walls of the cylinder. High $p$, $q$ orders are directly connected to the vortices introduced by the rotations of the impellers  and whose appearance is explainable in terms of  Goertler instabilities \cite{ravelet2008supercritical}. We will see in the next section that these effects are recovered also for global observables. \\

\subsubsection{Distance from Kolmogorov model}
The difference between the ($-$) and ($+$)  configuration is also highlighted by the results of the $\Upsilon$  computations reported in the middle right panels of Fig. \ref{armacon} and Fig. \ref{armaanti}. 
The distance from Kolmogorov model is lowest and almost zero near the boundaries, where the fluctuations are modest, and increases towards the center. 

Evident differences appear if one compares $\Upsilon$ values for ($+$) and ($-$) senses of rotation. As expected, the highest values are found in the ($+$) case, which is the one preserving a spatial four cells pattern in the fluctuations.
This suggests that the coherent structures visualized using bubble air seedings are responsible for   deviations from the Obukhov model. 

In the ($-$) set-up, the region of  values of $\Upsilon\geq 0.1$  clearly traces the area with maximal azimuthal velocities. We have therefore a clear connection between coherent structures. In the present data set, we do not observe obvious signature of the influence of the shear layer dynamics. However, by using a much larger data set, we have been able to evidence the signature of the wandering of the shear layer in between to metastable position. This is reported elsewhere \cite{faranda2014probing}.

\subsubsection{Physical interpretation of ARMA($p,q$) coefficients}
The bottom panels refer to the sum of the coefficients $\Phi=\sum_{i=1}^p | \phi_i |$  (bottom left panels) and $\Theta=\sum_{i=1}^q | \vartheta_i |$ (bottom right panels). $\Phi$ and $\Theta$ may be regarded as a representation of the total persistence of the phenomena i.e. how much the system remembers of its past history. 
In order to get a better understanding of this idea and thus obtain a physical interpretation of the AR and MA parts of the process, we exploit, once again, Thomson's model.  
Eq. \ref{T87} implies a Markovian evolution of the Lagrangian velocity in the inertial subrange, which is linked to an exponential behavior of the ACF: 
$$\rho(t) \sim e^{-\frac{t}{T_L}}.$$
We can observe that the first term of the r.h.s. of the equation contains some information about the global correlation structure of the process, which is even more evident in the discrete time ( see eq. (\ref{langAR})), since the autoregressive coefficient $\phi =  \left(1-\frac{\Delta t}{T_L}\right) $ is the Taylor expansion of the exponential ACF. We have already mentioned that $|\phi|$ is a measure of the persistence of the process; here, this persistence is driven by the large eddies, since $T_L$ is the Lagrangian decorrelation time scale, which is $T_L \gg \Delta t$, with $\Delta t$ lying in the inertial subrange.

On the other hand, since the noise term in eq. (\ref{T87}) is a standard Brownian motion, the innovations of the time series in eq. (\ref{langAR}) $\Delta W$, are normally identically distributed and $\delta$-correlated. Thus, the second term of the equation is simply a noise driving the process, with no linear dependence between any couple of values $(\Delta W_i, \Delta W_j)$ with $i \neq j$; this assumption is due to the fact that the stochastic kicks come from the viscous eddies, which live in the viscous time scale $\tau_{\eta} \ll \Delta t$: since the particle samples the turbulent field with a frequency linked to a characteristic time $\sim \Delta t$, the viscous eddies are completely uncorrelated between two steps. This means that here $q=0$ and the innovations of the process are a pure (Gaussian) noise, and all the information is contained in the mean and the variance.

In general, we may say that the AR($p$) part of an ARMA($p,q$) process is linked to the contribution of the large scales and represents the persistence of the process. Notice that, if $p=1$, it must be $|\phi| < 1$ in order to satisfy the stationarity condition; if the process is more persistent than an AR(1) with  $|\phi| < 1$, higher values of $p$ are required to explain all the correlation coming from the large scales. 
Analogous considerations hold for the MA($q$) part: if $q>0$, a linear combination of previous values of the noise appears in the equation, introducing a correlation structure in the innovation term, i.e. a higher persistence of the noisy contributions. This means that the small eddies do not decorrelate completely between two \textit{sampling times}, so we should assume to have eddies at all scales.

\subsubsection{Comparison with high-resolution datasets (LDV)}

The results obtained with the PIV technique must be validated and checked against higher  temporal resolution datasets. In fact, although the possibility of defining a distance from the Kolmogorov model in the physical  space rather than in the Fourier space  seems appealing, we must be sure that the results obtained with the ARMA analysis are stable with respect to an increase of resolution.\\
In the previous discussion, we have pointed out that the Obukhov model, representing homogeneous and isotropic turbulence, can be written as an ARMA(1,0) model. This corresponds to have a purely power-law spectrum which does not contain any other features than the decay predicted by Kolmogorov.  Since eq. (\ref{langAR}) contains an explicit dependence on   $\Delta t$ only for the coefficient $\phi$, we do not expect to see a change in the order of the process when increasing the resolution, but rather changes in $\phi$ and $\vartheta$ coefficients. This consideration holds unless the spectrum changes slope or has peaks  for some of the frequencies we add to the spectrum by increasing the resolution. In this case we expect to see also a change of the autoregressive and moving average polynomials.\\ 
In our analysis, we compared the PIV data for the ($-$) sense of rotation with an LDV experiment performed in the same conditions. Since for the LDV series only  the $w$ component is measured, we will compare this quantity to the same recorded  for the PIV experiments. The LDV data allow for exploring frequencies of order of the kHz, whereas the PIV is limited to a frequency of 15 Hz, so that we extend significantly the range of frequencies analyzed.   Whereas the time resolution of the LDV data is very high, the spatial resolution is  indeed low: we have $w$ measurements only at  the 18 points represented by  the red crosses in  the top-left panel of Fig.~\ref{LDV}. For this reason, the quantities obtained from the LDV analysis (left panels of Fig.~\ref{LDV}) have been interpolated on a  finer spatial grid. Anyway, the level of details remains lower  if compared with the  PIV results (right panels of Fig.~\ref{LDV}). The top panels of Fig. \ref{LDV} show a comparison between the averaged velocity field as obtained from the LDV and the PIV analysis. They both show not only the familiar cells structure, but also that the order of magnitude of the velocity fields is extremely close for the two different techniques. The analysis of the orders $\mathcal{O}$ (reported in the central panels of Fig.~\ref{LDV}) shows consistency between the two techniques:  the highest orders are located at the walls of the cylinder and near the impellers. Moreover, if we average the total order on all the available points, we get $\mathcal{O}=2.4\pm0.7$ for the LDV and $\mathcal{O}=1.8\pm 0.8$, values compatible within a standard deviation.  Finally, the analysis of $\Upsilon$ (lower panels of Fig.~\ref{LDV}) reveals that the maxima are located, both for the PIV and the LDV, near the walls of the cylinder around $Z=0$. By computing the average of $\Upsilon$ over all the points we find  $0.03\pm0.02$ for the LDV data and $0.02\pm 0.02$ for the PIV, again consistent within a standard deviation. 

\subsection{ARMA analysis  of phase randomized data: phase intermittency}

Phase randomization is often used in turbulence to destroy the intermittency effects related  to Fourier phases while preserving the intermittency effect related to Fourier amplitude \cite{nikora2001intermittency}: by applying such procedure one preserves up to the second order statistics (covariance and spectrum). It is therefore interesting to apply the ARMA analysis to phase randomized data sets to quantify the relative influence of phase and amplitude intermittency in turbulence.

A simple and efficient way to perform this phase randomization is to compute the Fourier transform of the time series, then randomize the phase (while preserving the anti-symmetry of the phase with respect to the frequency variable resulting from the real nature of the data) and going back to the physical space by means of an inverse Fourier transform. In order to perform this task, we have used the MATLAB code provided by Carlos Gias, based on the procedure described in \cite{prichard1994generating}.

The results we present correspond to the ($-$) sense of rotation for the PIV data already analysed in the previous sections. After generating surrogate velocity data, we compute $\Upsilon$ for the phase-randomized data and compare it with the original one. This is done in the lower panel of Fig. \ref{randophase}, with the original $\Upsilon$ (left) and the $\Upsilon$ for phase-randomized data (right). Both panels show the same structure meaning that most of the contribution to the intermittency
parameter is associated to intermittency amplitude through first and second order statistics (presumably through the advection and shearing effect of the large scale flow). To get information about phase intermittency, we subtract the intermittency index from the phase-randomized data to the original ones, obtaining $\Delta \Upsilon$. Results of such a difference are reported in the Upper panel of Fig. \ref{randophase}. It is about two orders of magnitude smaller than the amplitude intermittency. Its resulting spatial structure is highly organized, showing some association with  vortices. This feature may be connected with the observation that anomalous
scaling, in linearly advected hydrodynamical models, is
connected to the existence of statistically preserved structures
with highly complex geometrical properties \cite{biferale2002anomalous, celani2001statistical}.
 We leave this for future investigation.

\subsection{Hurst exponents}

A generalized version of  the first equation in the system \ref{T87} can be written as:

\begin{equation}
\diff u = a(x,u,t) \diff t + b(x,u,t)\diff W^{2H} \\
\label{langeH}
\end{equation}

where $\diff W^H$ is the increment of a fractional Brownian motion (fBm) and $H$ is the so called Hurst exponent \cite{beran1994statistics}. 
The fBm, firstly introduced by \cite{mandelbrot1968fractional},  is a generalization of  Brownian motion where the   increments are not independent. It has  zero mean and the following covariance function:

$$E[W_H(t) W_H (s)]=\frac{1}{2} (|t|^{2H}+|s|^{2H}-|t-s|^{2H})$$.

The exponent $H$ is a real number in $(0, 1)$ and its value determines  the memory of the stochastic process. For $H=1/2$, the standard Brownian motion is recovered. For $0<H<1/2$ the process is anti-persistent, i.e. an increase will most likely be followed by a decrease or vice-versa. Finally, for  $1/2<H<1$, the series is persistent, i.e.  increases generally follow increases. 

We want to investigate if the behavior displayed by the total order and total persistence of the ARMA($p$,$q$) and by the distance index $\Upsilon$ can be better explained by the fractional nature of the underlying stochastic process. In order to do this, we compute $H$ for  the $|\vec{V}(t)|$ data in the  ($+$) sense of rotation and we compare it to $\Upsilon$ in Fig. \ref{hurst}. The computation of the Hurst exponents follow the methods presented in \cite{weron2002estimating}  which we found to be all consistent with each other. Since $\Upsilon$ values span 5 orders of magnitude  while $H$ is always of order 1,  we consider the  $\log_{10}(\Upsilon)$. Not only the spatial structure of $H$ and $\log_{10}(\Upsilon)$ are very similar (upper panels of Fig.~\ref{hurst}),  but also a linear relation can be found between these two quantities (lower panel of the same figure). The linear correlation coefficient is $r=0.70$ and these results hold also for the LDV experiments with almost identical fit coefficients and $r=0.81$. From this analysis we argue that a fBm description of the phenomenon might be used to explain the nature of the correlations in the series and it could be useful to further improve the modeling of inhomogeneous and anisotropic turbulence. However, the results show that $\Upsilon$, based only on an ARMA($p$,$q$) estimation, is equally effective in quantifying deviations from the Kolmogorov model which could not be due to fBm effects.

\subsection{Global observables}

An interesting question to address when dealing with spatial-temporal extended systems, is how the information on the single trajectories is transmitted to integrated quantities.  In particular, one may ask whether the  differences found in the ARMA analysis for the local observables of PIV fields are preserved for scalar quantities i.e. if high ARMA($p,q$) orders found locally in the proximity of the impellers and the cylinder walls give a contribution to global observables or whether they average out. In this section we present  results obtained for the quantity $\delta(t)$ introduced in eq. (\ref{delta}). We have further tested that our results are independent of the choice of the global observable, whether derived from PIV measurements - Angular momentum - or  measured independently like  torques measurements.\\

\begin{figure}
\includegraphics[width=0.5\textwidth]{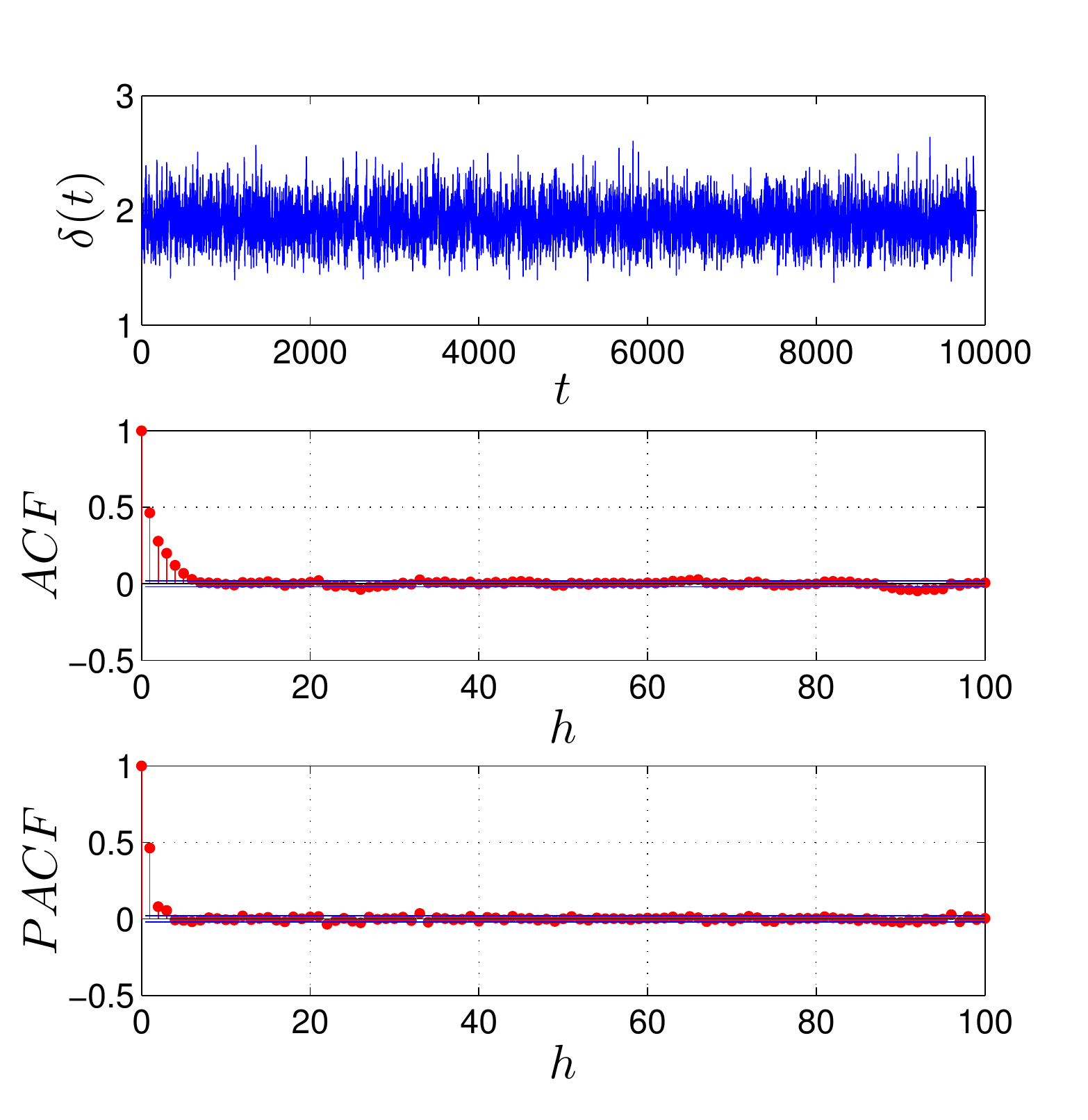}\includegraphics[width=0.5\textwidth]{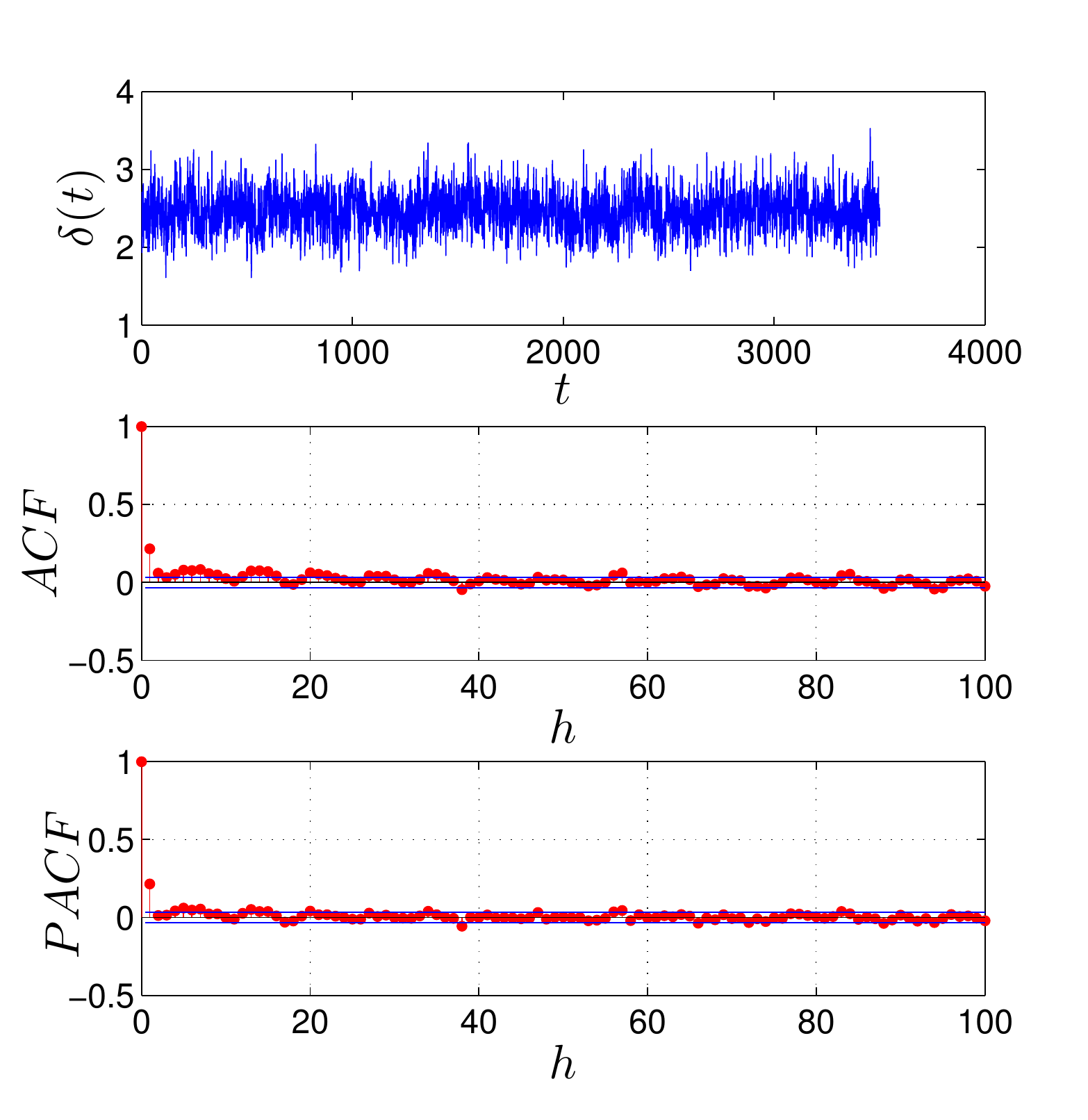}
\caption{Time series of $\delta(t)$ (upper panels) with their respective ACF functions (middle panels) and PACF (lower panels). Left: ($+$) sense of rotation. Right: ($-$) sense of rotation. $Re\simeq10^5$.  Blue lines in the ACFs and PACFs represent the confidence bands at the 95$\%$ confidence level. Sample frequency: 15 Hz. X-axis is in sample index.} 
\label{armadelta}
\end{figure}

\begin{figure}
\includegraphics[width=0.8\textwidth]{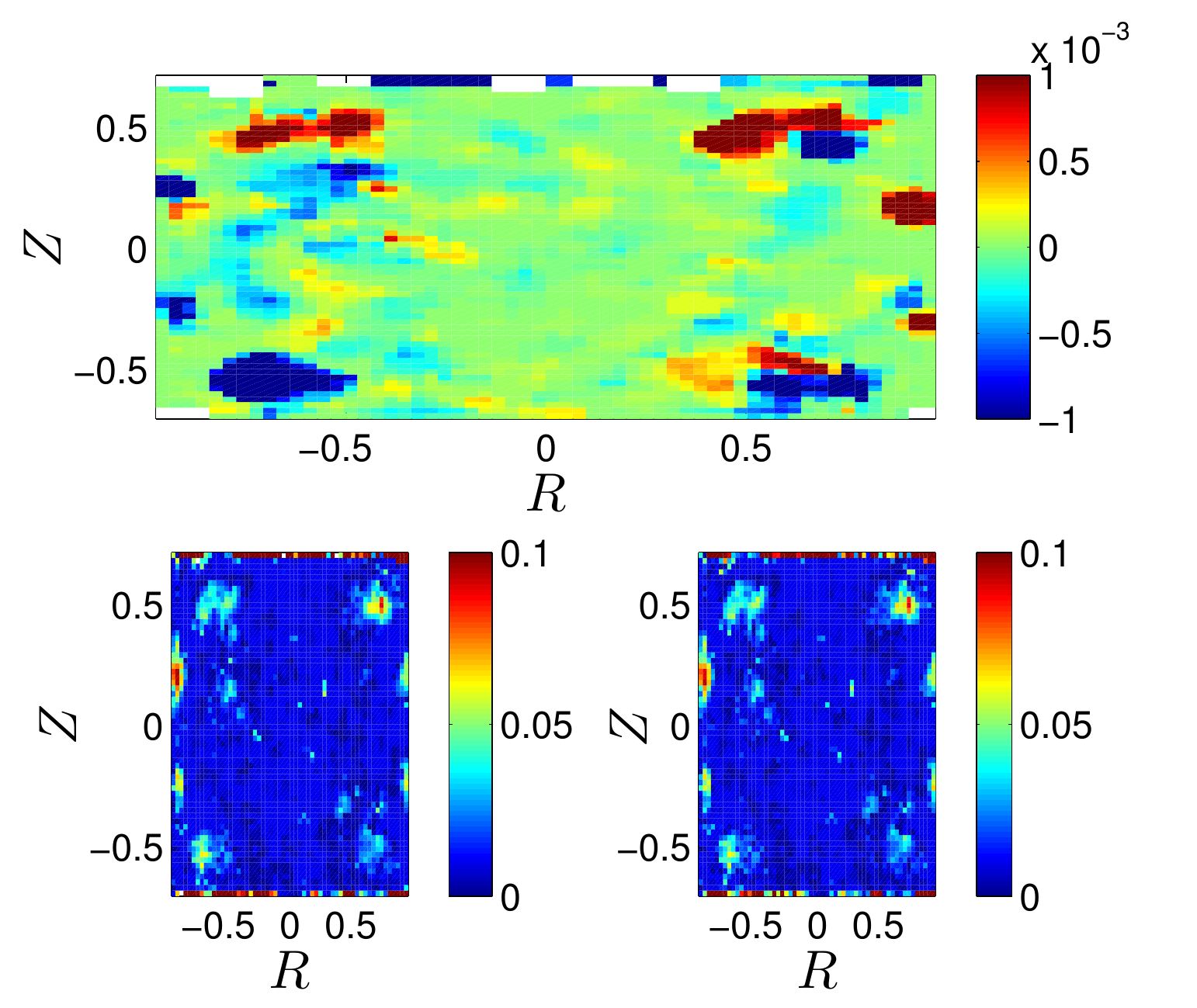}
\caption{Upper panel: Difference between $\Upsilon$ computed for the $|\vec{V}(t)|$ data for the  ($-$) sense of rotation and $\Upsilon$ on surrogate date of the same experiment obtained with a phase randomization procedure.  Lower panel: $\Upsilon$ for the original data (left) and for the surrogates (right).}
\label{randophase} 
\end{figure}

\begin{figure}
\includegraphics[width=0.5\textwidth]{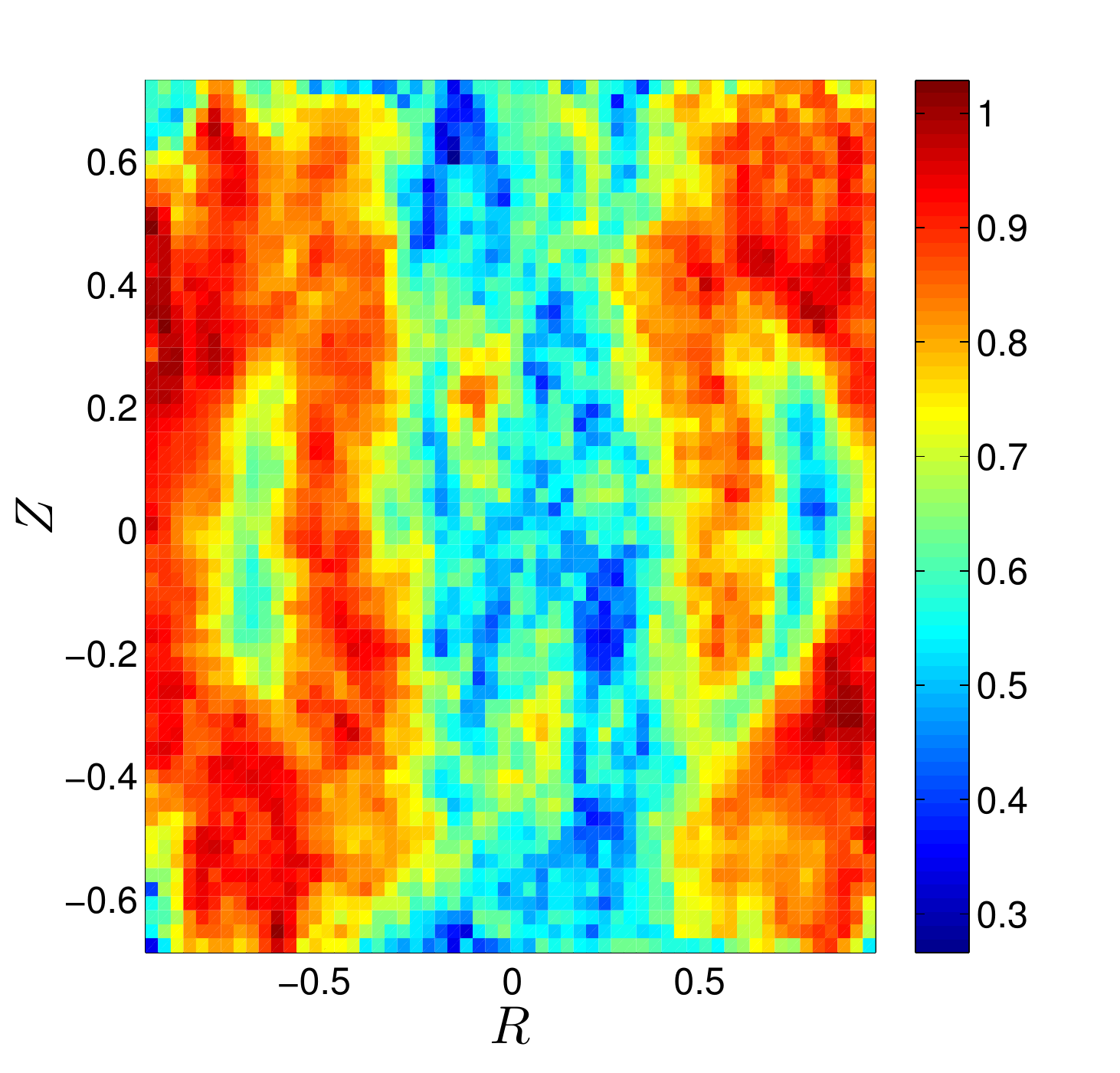}\includegraphics[width=0.5\textwidth]{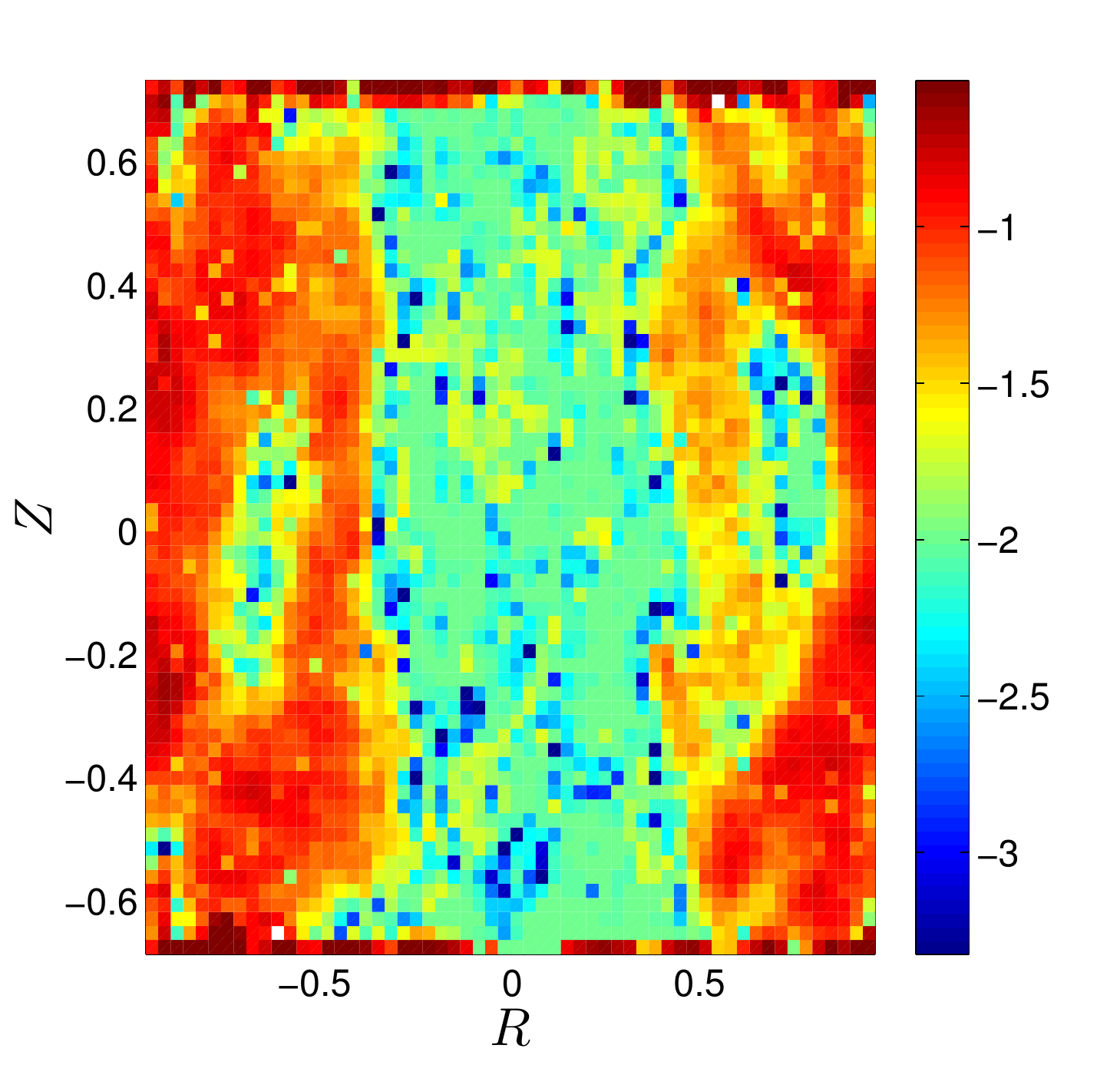}\\
\includegraphics[width=0.7\textwidth]{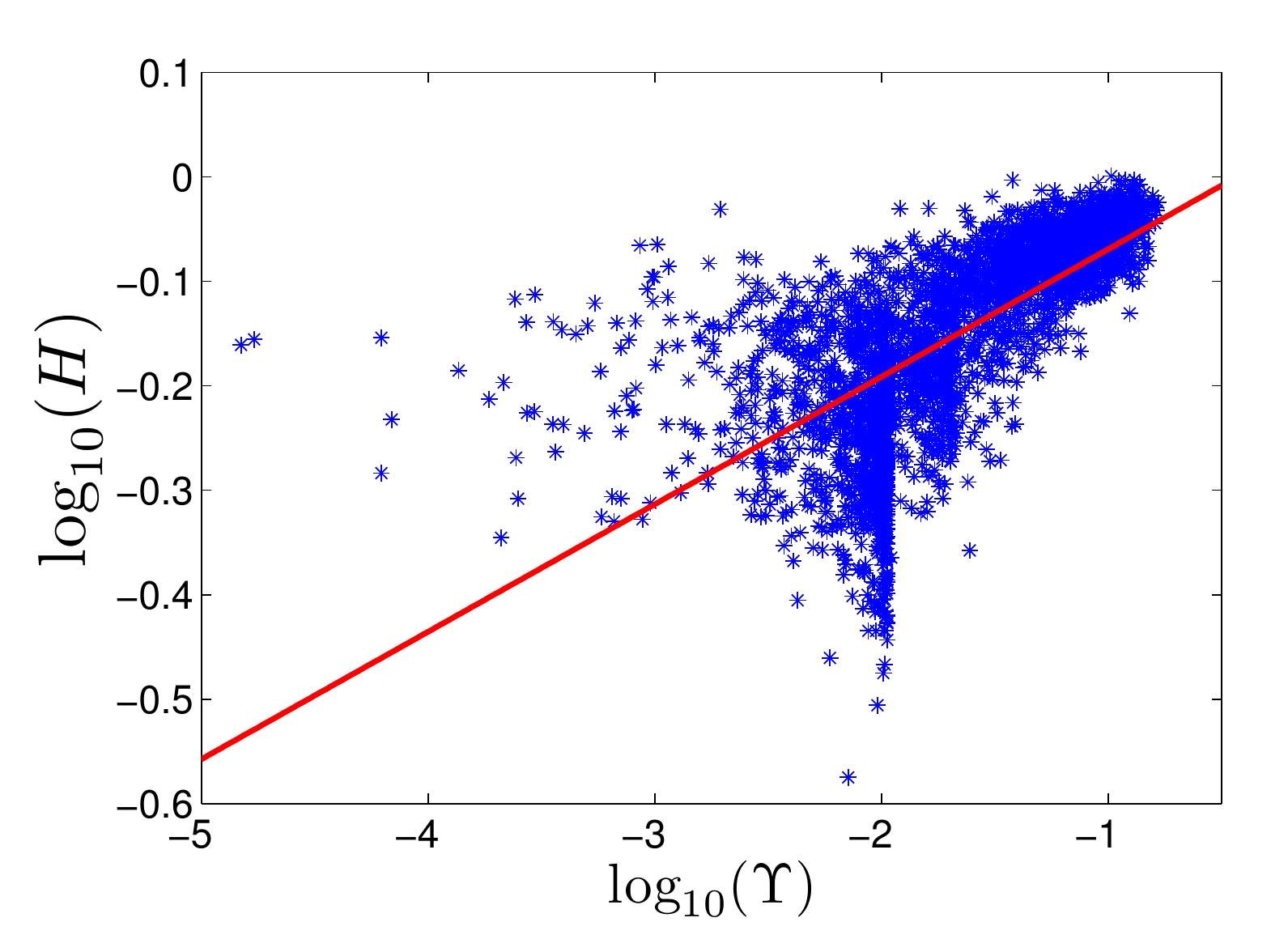}
\caption{Upper panels: comparison between the Hurst exponents $H$ (left) and the deviation from the Kolmogorov model in log-scale $\log_{10}(\Upsilon)$ (right)  found  by fitting an ARMA($p$,$q$) to the $|\vec{V}(t)|$ data for the  ($+$) sense of rotation. Lower panel: scatter plot of the Hurst Exponent $H$ and $\log_{10}(\Upsilon)$ . The red solid line  shows a linear fit to the data. }
\label{hurst} 
\end{figure}

\begin{figure}
\includegraphics[width=1\textwidth]{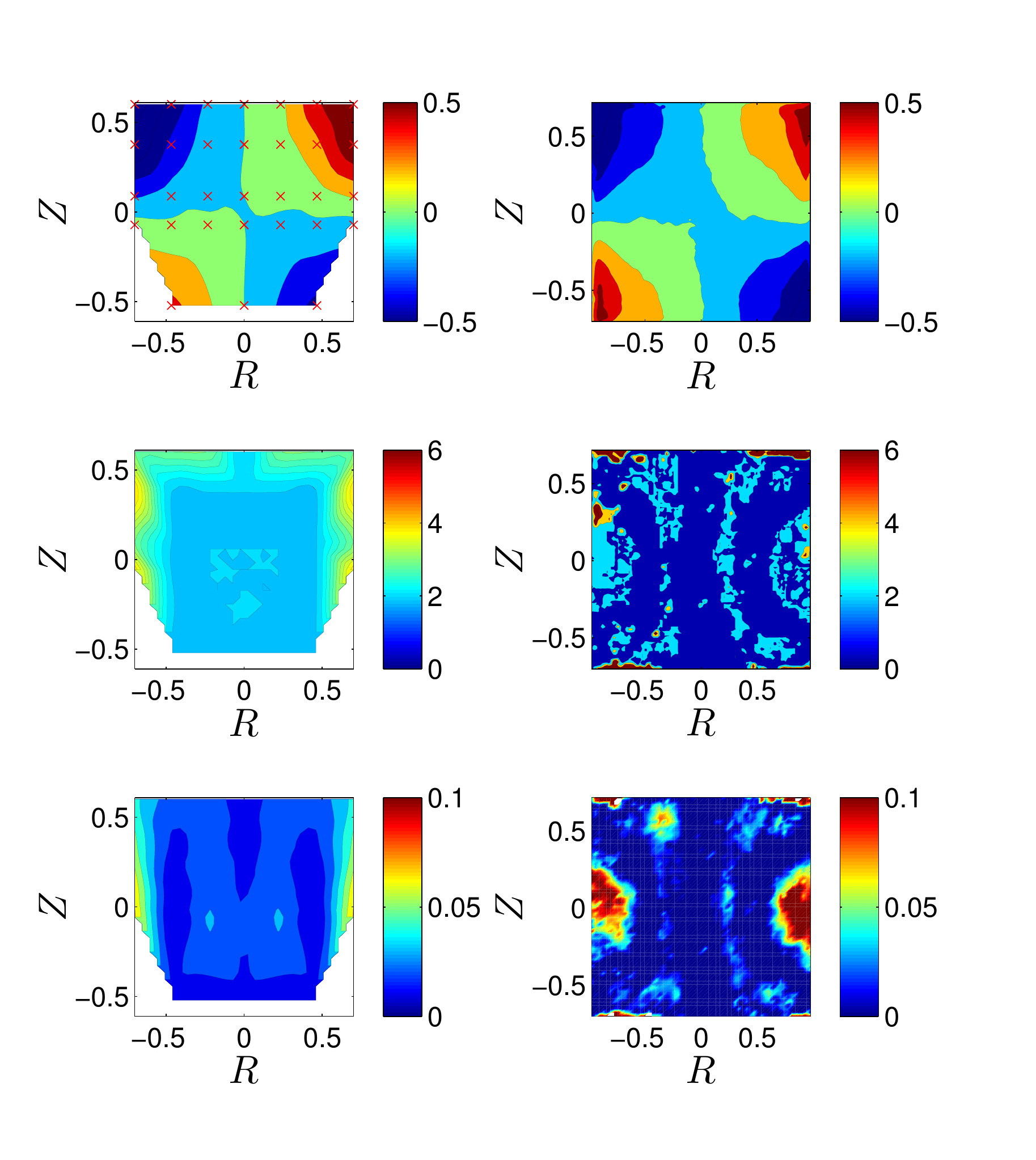}
\caption{Comparison between the ARMA analysis for LDV data (left) and PIV data (right) for the azimuthal component of the velocity $w$.  $\bar{w}$ (upper panels), total order $\mathcal{O}$   (central panels) and   distance from Kolmogorov model $\Upsilon$ (lower panels); ($-$) sense of rotation; $Re\simeq10^5$. The red crosses in the top-left panel show the locations of the measurement points for the LDV experiment.} 
\label{LDV}
\end{figure}

We have carried out the analysis on global observables at several Reynolds numbers around $Re=10^5$, that is in a fully turbulent regime.  The typical behavior of the time series of $\delta(t)$  is represented in Fig. \ref{armadelta} for the ($+$) sense of rotation (left), and the ($-$) one  (right). The top panel refers to the time series obtained by averaging the spatial velocity fields and shows no particular differences at first sight, as we have seen for the examples of the velocity series shown in Fig. \ref{armaex}. However, the ACF (middle panels) and PACF (lower panels) are remarkably different. The ACF of the ($+$) sense of rotation decays quickly and the PACF shows only one peak significantly different from zero: an ARMA(1,0)  is enough to explain the correlation structure. On the other hand, an oscillatory behavior of both the ACF and PACF is clearly recognizable for the ($-$) rotation. The orders $p,q$ needed to decorrelate the latter time series are higher in the ($-$) rotation, namely $p=2$, $q=1$. These results hold generally by varying  $Re$ and changing observables and point to the intrinsic differences between the two senses of rotation.

One can notice that some characteristic features appear in the ACF of the ($-$) sense of rotation. We can speculate that, for scalar quantities, the highest orders get averaged out if their contribution is substantially different at different $(r,z)$ as it happens for the ($+$) rotation. However, when the same kind of features are present in the ACF and PACF for series at different $(r,z)$ the contribution sums up and is well visible in the behavior of global observables.\\

\section{Multistability}

\begin{figure}
\includegraphics[width=0.7\textwidth]{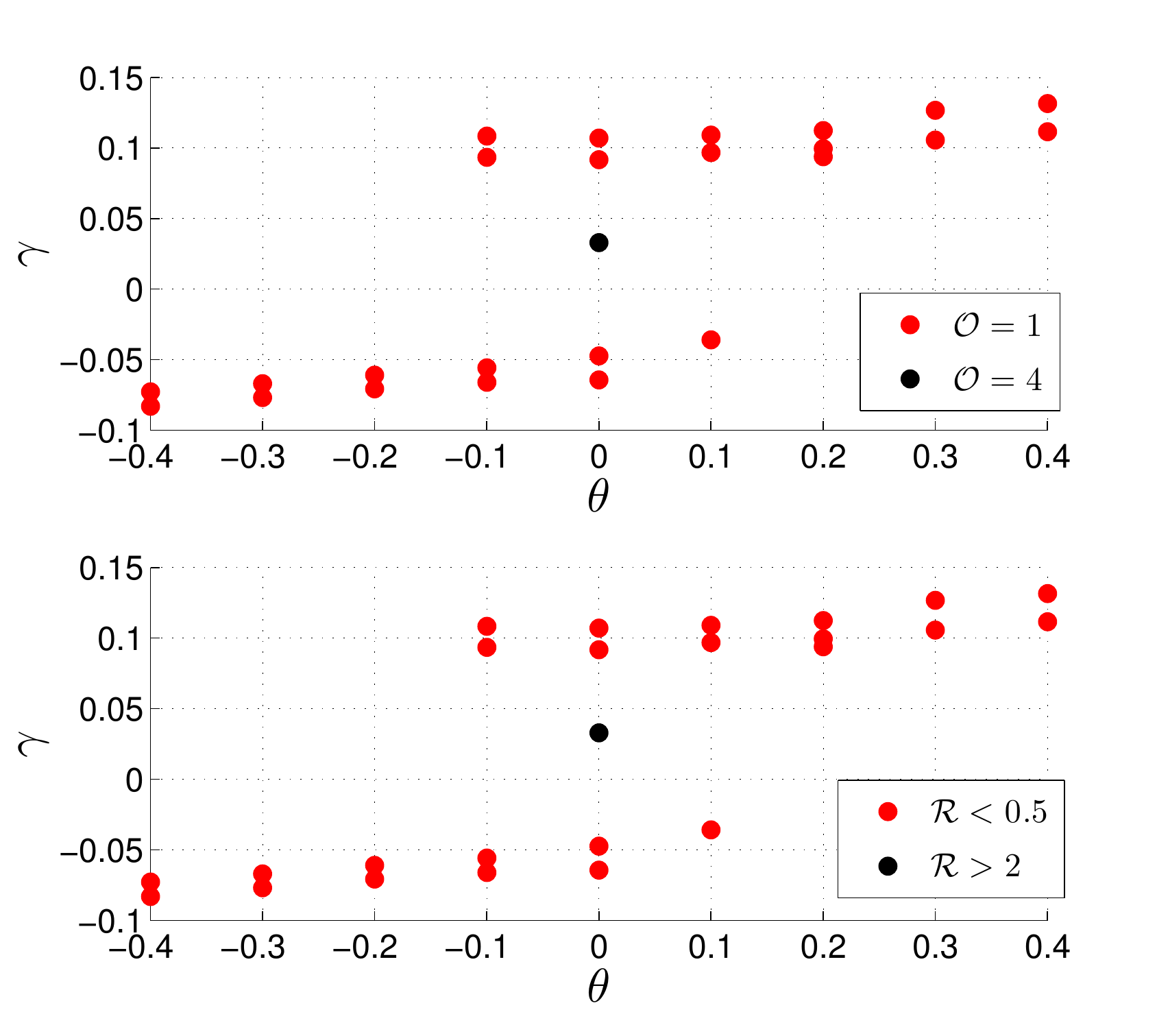}
\caption{ total order $\mathcal{O}$ and total persistence $\mathcal{R}$ for the von K\'arm\'an experiment under the speed control. The points represent the averaged $\gamma$ and $\theta$ obtained for each experiment}
\label{vitesse}
\includegraphics[width=0.7\textwidth]{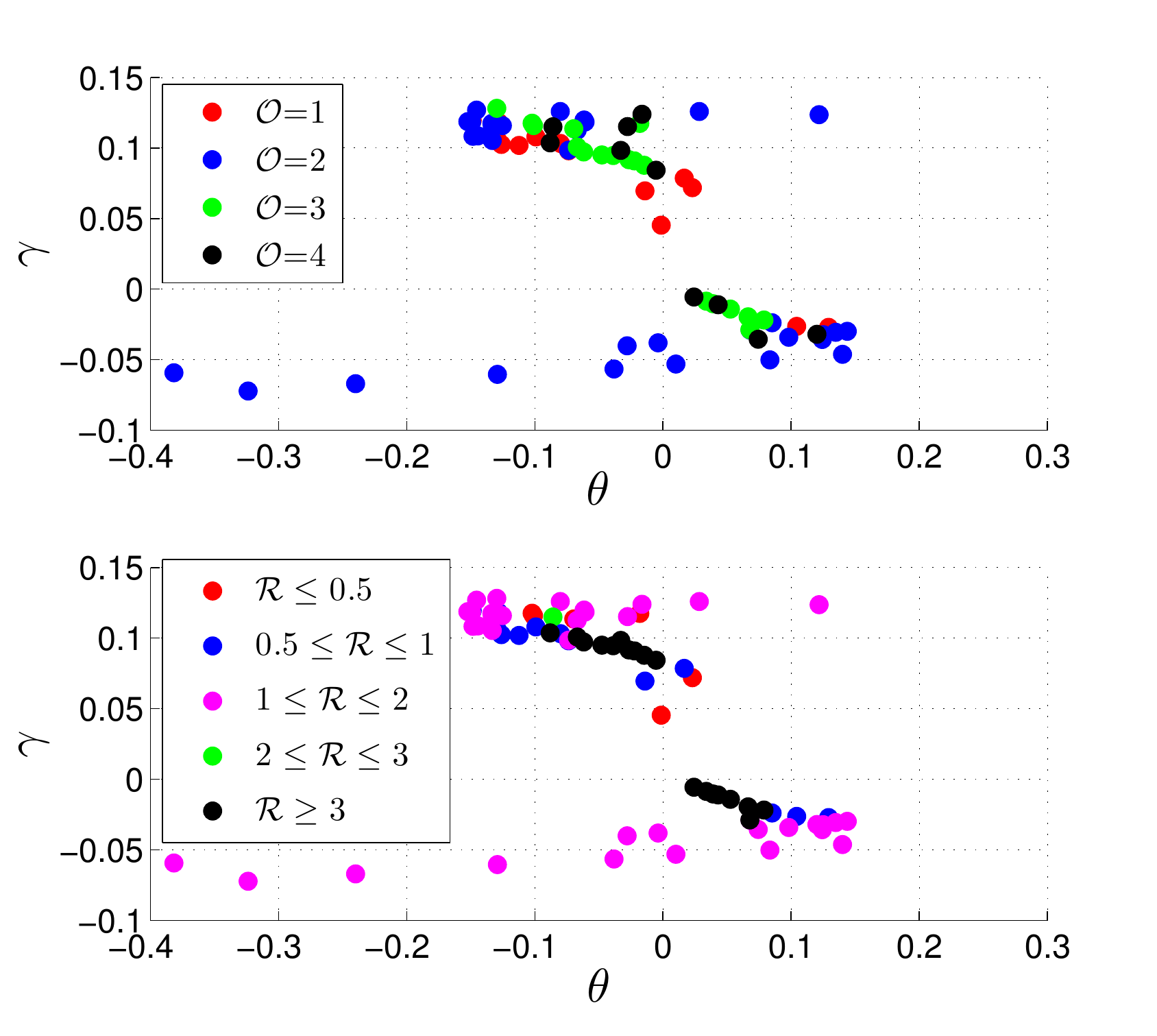}
\caption{ total order $\mathcal{O}$ and  total persistence $\mathcal{R}$ for the von K\'arm\'an experiment under the torque control. The points represent the averaged $\gamma$ and $\theta$ obtained for each experiment} 
\label{couple} 
\end{figure}

 Another interesting question is whether  the application of ARMA techniques to turbulence  is helpful to discriminate between different stability regimes. The simple guess is that by increasing the instability of a configuration, higher orders  $\mathcal{O}$ arise as we introduce  in the system new time scales  linked to the presence of nearby attracting states. In order to check this idea, let us consider again a  von K\'arm\'an swirling flow with the same  geometry  described before, with   the Reynolds number   fixed  at $Re \sim 10^5$.  In this system, one can impose either the speeds $f_1$ and $f_2$ of the motors or the torques $C_1, C_2$ and define two natural dimensionless quantities:

$$\theta=(f_1-f_2)/(f_1+f_2), \quad \gamma=(C_1-C_2)/(C_1+C_2)$$

which are respectevely the reduced impeller speed difference and the reduced shaft torque difference.  In \cite{briceprl}, the authors found that different forcing conditions change the nature of the stability of the steady states. Here we  complete  the results  represented in Fig. 1 of  \cite{briceprl}, with the ARMA analysis in terms of the quantities $\mathcal{O}$ and the total persistence of the process defined as $$\mathcal{R}= \sum_{i=1}^p|\phi_i| +\sum_{i=1}^q|\theta_i|.$$\\
\textit{Speed control.}  In this case all the turbulent flows are steady.  By plotting  the averaged $\theta$'s and $\gamma$'s  measured in several experiments, one  obtains the diagram shown in Fig. \ref{vitesse}. The colors refer  to different $\mathcal{O}$ (Fig \ref{vitesse}-upper panel), and $\mathcal{R}$ (Fig \ref{vitesse}-lower panel). Starting both impellers at $\theta\simeq 0$ leads to a marginally stable state, which consists of two symmetric recirculation cells separated by a shear layer.   If one waits enough time, a fluctuation may force a jump of the system to one of the two bifurcated states represented by the red points. The instability of the symmetric state is reflected by the order of the ARMA processes fitted for the series of $\gamma(t)$ at $\gamma\simeq \theta \simeq 0$.  For this experiment we found $\mathcal{O}=4$ and   $\mathcal{R}\simeq 3$, values  definetely larger than the ones found in the bifurcated states  where  always  $\mathcal{O}=1$ $\mathcal{R} < 0.5$. From the available data one can argue that  the potential barrier - the repellor in dynamical system  - at $\theta=\pm 0.1$ is somehow impenetrable as we do not get any increase in  $\mathcal{O}$ and $\mathcal{R}$ for the series $\gamma(t)$ recorded at such values of $\theta$.\\
\textit{Torque control.}  By imposing the torque control one   gets access to new attracting states, located in correspondence to the repellor found in the speed control. The results for the torque control have been obtained by analysing time series of $\theta(t)$ and the results in terms of $\mathcal{O}$ and $\mathcal{R}$  are reported in Fig. \ref{couple}.  Before commenting on the new states, we begin by analysing the states which are attracting in both the configurations.   In \cite{briceprl}, the authors assert that the properties of the attracting states in the speed control set up are analogous to the ones found for the torque control. However, by applying the ARMA analysis, we found, as one would expect, remarkable differences.  In particular, the symmetric state  (which was marginally stable in the speed control) is now stable as one can go to the bifurcated states in a continuous way.  This is confirmed by the low values of  $\mathcal{O}$ and $\mathcal{R}$  found for the torque control  where $\mathcal{O}=1$ (Fig~\ref{couple}-a, found in correspondence of $\gamma\simeq0.005$ and $\theta\simeq0$) and $\mathcal{R} <1$ (Fig~\ref{couple}-b).  The bifurcated states have a different characteristic order (typically $\mathcal{O}=2$) and persistence   (typically $1<\mathcal{R}<2$).  These are not linked to the presence of transitions, as they persist further away from the unstable range of parameters. They are linked to the modifications in the dynamics induced by the change of control which  affects  also the stable regions in a fine way, which has not been discussed in  \cite{briceprl} but is evident by applying the ARMA technique capabilities.
Let us now comment on the new states which appear in the torque control in correspondence to the repellor for the speed control. These states feature multistability as detailed in \cite{briceprl}. In terms of ARMA analysis they are characterized by higher orders  (green branches in the upper panel of  Fig \ref{couple} with  $\mathcal{O}=3$), and persistence (black branches of the lower panel in Fig \ref{couple}). Even if  an order is found for multistable time series, an ARMA model cannot just be sticked to the data in this case, as it will not reproduce a multistable behavior, but rather a process with one stable state whose correlation properties are similar to the ones found for the multistable time series.\\
This example clearly shows that one can find, far from the bifurcation, a typical ARMA($p$,$q$) process ($\mathcal{O}=1$ for the speed control and   $\mathcal{O}=2 $  for the torque control)  which describe the data-sets whereas $p$ and $q$ are evidently modified by the stability properties.  We remark that one must pay extraordinary care when the goal is to find a model for the data-set. \\

\section{Final remarks}


In this paper we have shown how to extract information from turbulent data-sets by applying an ARMA statistical analysis. Such analysis goes well beyond the analysis of the mean flow and of the fluctuations; in fact, it is possible to link the behavior of the recorded time series to a discrete version of a stochastic differential equation which is able to describe the correlation structure in the data-set.
We have tested the method on  data-sets produced by the experiments on the von-K\'arm\'an swirling flow, already analyzed in several publications. \\

We have shown how the anisotropies and inhomogeneities present in real experiments as well as the finite resolution of the data-sets influence the order $p$, $q$ of the process which better describes the data. We find that data are suitably described by ARMA($p,q$) processes whose orders are generally different from the Obukhov model although with a very limited number of auto-regressive and moving average terms (generally $p, q=1$ or 2). We have introduced a new index $0\le \Upsilon\le 1$ to measure and quantify this difference. The value of $\Upsilon$ increases in areas where large scale coherent structures are present. It would be interesting to rely the statistics  of $\Upsilon$  to the computation of refined statistics of velocity increments, possible only for time high-resolved experiments \cite{arneodo1996structure}.  In particular,  we aim to compare $\Upsilon$ with classical intermittency parameters based on structure functions.  This idea follows from the hypothesis firstly proposed by \cite{laval2001nonlocality} that intermittency \textit{propagates} in direct interactions between large and small scales, rather  than in cascades.
Preliminary analysis carried on the LDV data-sets show that there is a linear proportionality between $\Upsilon$ and the classical intermittency parameters defined on the velocity increments. \\

The inhomogeneous structure of the PIV experiments is reflected by the range of different orders $p$, $q$ found in our analysis:  a great part of the flow can be described in terms of noise, whereas higher orders concentrate around cells (($+$) sense of rotation) or near the walls (($-$) rotation).  The analysis of global observables  shows that most of the information about the local structure of the flow is preserved, including the differences found between the two senses of rotation. This correspondence between local and global quantities is very important and it will be further exploited for challenging systems for which only global observables are available as in the SHREK  data-sets experiment with super-fluid Helium \cite{collaboration2014probing}.\\ 

We have also checked our results against the change in  time resolution by comparing them against the LDV experiments, whose average sampling frequency is order of the kHz. Not only the values of $\mathcal{O}$ found with this technique are consistent, but also the values of $\Upsilon$ and the spatial structures observed. Finally, we have commented on the effects of multistability on the ARMA analysis by considering two different kind of forcings for the von K\'arm\'an experiment.

In the Obukhov model, the coefficients of the stochastic model  are given from the turbulence theory, resulting in a simple Langevin equation which describes the process in the continuous time. Here, we have applied estimation methods to obtain a parametric description in the discrete time, but the passage to continuous time stochastic differential equations is not trivial for a general ARMA($p,q$) process. Obtaining an expression of the model's parameters in terms of physical quantities of turbulence theory is presently not possible.
In fact, it is likely that the MA($q$) part of our processes represents the contribution of the shortest time scales detectable with the available techniques.  One way to test this idea is to verify that only the structure of the MA($q$) part of the process changes by changing only small scales feature of the flow. In order to do that, we are currently testing  impellers with a fractal structure. Preliminary analysis show that  MA($q$) orders are different between fractal and non-fractal impellers whereas the AR($p$) do not change. Details will be reported in a future publication. \\

Several generalizations of ARMA models exist and they allow for taking into account  the possible multi-fractal behavior of turbulence. The comparison of the Hurst exponent and the $\Upsilon$ index suggests that it will be interesting to extend the analysis to fractional integrated ARMA or  ARFIMA($p$,$H$,$q$) models. The first ones can be appropriated for studying problems of non-stationary turbulence \cite{connaughton2003non} whereas  a SARMA models analysis could be suitable for studying problems of wave turbulence\cite{nazarenko2011wave}. 

\section{Acknowledgments}
We acknowledge the other members of the VKE collaboration who performed the experiments:  C Wiertel and V Padilla.  We acknowledge two anonymous referees whose comments greatly improved the results and the consistency of this paper. DF acknowledges the support of a CNRS postdoctoral grant.

\bibliographystyle{ieeetr} 
\bibliography{ARMAbib}

\section{Appendix: A theoretical survey on ARMA modeling}

We have already mentioned, in the definition of the ARMA($p,q$) process, that $\{ X_t \}$ must be stationary. Usually, two definitions of stationarity are given when treating discrete time stochastic processes: {\itshape strong stationarity} implies stationarity of the whole joint probability distribution of the stochastic process, while {\itshape weak stationarity} requires the first two moments of the process to be finite and constant in time. The results about ARMA($p,q$) processes are usually proved requiring weak stationarity, that is of course implied by strong stationarity;  in our data analysis we will assume stationarity on a physical basis, by studying the system when the dynamics has reached well identifiable stationary states .

First of all, we observe that eq. (\ref{ARMA1}) can be written in a very compact form, introducing the backward operator $B$ such that $B^j X_t = X_{t-j}, j \in \mathbb{Z}$:
\begin{equation}\label{ARMA2}
\phi(B)X_t = \vartheta(B) \varepsilon_t.
\end{equation}
If $\vartheta(z) \equiv 1$ the process reduces to an AR$($p$)$, if $\phi(z) \equiv 1$ the process is said to be a MA$(q)$ process.

Formally, existence and uniqueness of a stationary solution $\{ X_t \}$ of eq. (\ref{ARMA2}) are satisfied if and only if:
\begin{equation}\label{station}
\phi(z) \neq 0 \hspace{5mm} \forall \, |z| = 1.
\end{equation}

Two other important features of a discrete time stochastic process are causality and invertibility. Causality refers to the possibility of recovering the value of the process at time $t$ as a function of the innovations $\varepsilon_s$, with $s \leq t$. Formally, $\{X_t \}$ is causal if there exists a succession of absolutely summable coefficients $\{\psi_j \}$ so that the process at time t can be written:
\begin{equation}\label{caus1}
X_t = \sum_{j=0}^{\infty} \psi_j \varepsilon_{t-j}
\end{equation}
which implies, in terms of the auto-regressive polynomial:
\begin{equation}\label{caus2}
\phi(z) \neq 0 \hspace{5mm} \forall \, |z| \leq 1. 
\end{equation}
Invertibility could be regarded as the property specular to causality, so $\{X_t \}$ is invertible if the series of the innovations $\{\varepsilon_t \}$ can be recovered from the process. This requires the existence of a succession of summable coefficients $\{\pi_j \}$, which allows us to write:
\begin{equation}\label{inv1}
\varepsilon_t = \sum_{j=0}^{\infty} \pi_j X_{t-j}
\end{equation}
and in this case, condition \ref{caus2} is required on the moving-average polynomial:
\begin{equation}\label{inv2}
\vartheta(z) \neq 0 \hspace{5mm} \forall \, |z| \leq 1. 
\end{equation}

In case of the presence of $d$ unit roots in the auto-regressive polynomial, the process becomes non stationary; however, the $d$-differenced process $(1-B)^dX_t$ can be stationary. In particular, if $(1-B)^d X_t$ is an ARMA$(p,q)$ process, $X_t$ is said to be an ARIMA($p,d,q$) process (where the 'I' stands for {\itshape integrated}); the particular case of an AR(1) with $\phi=1$ reduces to the well-known random walk. Taking the differences of a time series is a drastic operation and a careful testing for the presence of unit roots must be performed if this kind of non stationarity (also called {\itshape stochastic trend}) is supposed to exist. The most used test to this purpose is the Augmented Dickey-Fuller test; notice that a unit root in the moving average can also be taken into account through the hypothesis that the innovations are an integrated autoregressive process.
Extensions to more complicated models can be found in literature, but these basics ARIMA processes are sufficient for the data analysis proposed in the present work.

The main idea of time series analysis through ARMA models is to select the linear model that fits the data in the most parsimonious way, so that diagnosis of the nature of the generating process, forecasting or Monte Carlo simulations can be performed. Model selection is a non-trivial step of the procedure and should be discussed after the introduction of some fundamental tools for the investigation of the time-dependence structure of the stochastic process. This issue can be addressed through spectral analysis, decomposition of the time series in trend, cyclical, periodical and irregular components and, most of all, correlation analysis. In this approach the dependence structure is studied analyzing the global and the partial autocorrelation functions. 

The (global) auto-covariance function (ACVF) at lag $h$ of a zero-mean stochastic process is defined: 
\begin{equation}\label{gacv}
\gamma(h) = E[X_{t+h} X_t]
\end{equation}
and, when normalized over the variance, gives the (global) autocorrelation function (ACF):
\begin{equation}\label{acf}
\rho(h) = \frac{\gamma(h)}{\gamma(0)}.
\end{equation}

The concept of partial autocorrelation function (PACF) is less intuitive; formally, it can be written:
\begin{eqnarray}\label{pacf}
\alpha(0) & = & 1 \\ \nonumber
\alpha(h) & = & \phi_{hh} \hspace{5 mm} h = 1, 2, \dots \\ \nonumber
\end{eqnarray}
where $\phi_{hh}$ is the last component of $\phi_h = \Gamma_h^{-1}\gamma_h$ with $\Gamma_h = [\gamma(i-j)]_{i,j=1}^h$ and $\gamma_h = [\gamma(1), \dots, \gamma(h)]'$.
In practice, the PACF quantifies the correlation between the prediction errors at lag 0 and $h$, given that it can be shown that the conditional expected value is the best linear predictor: $$\phi_{hh} = CORR[X_h - P(X_h | X_1, \dots, X_{h-1}), X_0 - P(X_0 | X_1,\dots, X_{h-1})].$$

For very simple processes, the ACF and the PACF give strong hints about the time dependence  structure. In particular, for a MA$(q)$ process, the theoretical ACF is characterized by $q$ non-zero peaks, while the PACF decays exponentially or as a damped trigonometric function; for AR($p$) processes specular considerations are valid. 
On the ($+$), if the process is characterized by both autoregressive and moving-average polynomials, few information can be obtained by a simple by-eye evaluation of the ACF and PACF. In this case, statistical information criteria must be used; one of the most widely known is the Bayesian Information Criterion ($BIC$), which will be presented after the introduction of the concept of estimation for ARMA$(p,q)$ models.

Given a parametric hypothesis ARMA$(p,q)$ for a time series, the corresponding discrete-time equation is fitted to the data and all the parameters are estimated with maximum likelihood techniques, well described also in the more practical volume \cite{brockwell_etal-springer-2002}. At this point, two sets $\{\hat{\phi}_j\}_{j=1}^p$ and $\{\hat{\vartheta}_j\}_{j=1}^q$ of estimated parameters are available, as well as a time series of estimated residuals $\{\hat{\varepsilon}\}$ of the same length of the original time series; if the tested ARMA$(p,q)$ fits the data, $\{\hat{\varepsilon}\}$ must be a sequence of independent random variables. Notice that, if the orders $p$ and $q$ are too high, the time series is over-fitted, so the analyst must be careful in choosing the most parsimonious model in terms of number of parameters. Thus, once the residual sequence has been obtained, inference must be made on the null hypothesis $H_0$ of uncorrelated residuals. At this point, the sample ACF $\hat{\rho}(j)$ is computed; then, one of the most used test statistics is the Ljung-Box Test:
\begin{equation}\label{LB}
Q_{LB} = n (n+2) \sum_{n-j}^h \frac{\hat{\rho}(j)^2}{n-j} \sim_{H_0} \chi^2(h)
\end{equation}
where $n$ is the length of the time series and $h$ a fixed number of lags at which the sample ACF is computed. If $H_0$ is not rejected at a given level (usually $\alpha=0.01$ or $\alpha = 0.05$), the tested ARMA$(p,q)$ fits the time series.

As already mentioned, in case of complex or high-order processes, ACF and PACF are not sufficient to obtain a hint on the possible order $(p,q)$ at glance; in this case, some different hypothetical ARMA$(p,q)$ models can be fitted and for each one the $BIC$ is computed:
\begin{eqnarray}\label{BIC}
BIC & = & (n-p-q)\ln \left[ \frac{n\hat{\sigma}^2}{n-p-q} \right] + n(1 + \ln\sqrt{2\pi}) +\\
       & + & (p+q)\ln\left[ \frac{\left(\sum_{t=1}^n X_t^2 -n\hat{\sigma}^2  \right)}{p+q} \right]. \nonumber
\end{eqnarray}
This quantity is minimized by the most parsimonious model providing a good fit to the time series. However, an other possibility is to fit many different ARMA$(p,q)$ models and choose the one for which the null hypothesis of uncorrelated residuals is not rejected and the total number $(p+q)$ of parameters is minimum. 
\newpage


\end{document}